%% file: 20211208_WedgeHybridEnergyFlux.tex
\documentclass{article}

\usepackage{PRIMEarxiv}

\usepackage[utf8]{inputenc} 
\usepackage[T1]{fontenc}    
\usepackage{hyperref}       
\usepackage{url}            
\usepackage{booktabs}       
\usepackage{amsfonts}       
\usepackage{nicefrac}       
\usepackage{microtype}      
\usepackage{fancyhdr}       
\usepackage{graphicx}       
\usepackage{amsmath}
\usepackage{amssymb}
\usepackage{mathrsfs}
\usepackage{bm}
\usepackage{units}
\usepackage{cancel}

\title{Hybrid normal mode and energy flux model for an ideal oceanic wedge environment with radial sound speed front}

\author{
  Mark Langhirt \\
  Penn State Graduate Program in Acoustics \\
  \And
  Charles Holland \\
  Portland State University \\
  \And
  Sheri Martinelli \\
  Penn State University \\
  \And
  Ying-Tsong Lin \\
  Woods Hole Oceanographic Institute \\
  \And
  Dan Brown \\
  Penn State University \\
}

\begin{document}

  \include{definitions}

  \maketitle

  \begin{abstract}
    Energy flux is an acoustic propagation model that calculates the locally-averaged intensity without computing explicit eigenvalues or tracing rays.  The energy flux method has so far only been used for two-dimensional problems that have collapsed the third dimension by rotational or translational symmetry. This report outlines the derivation and implementation of a three-dimensional ocean acoustic propagation model using a combination of normal modes and the energy flux method.  This model is specifically derived for a wedge environment with a radial sound speed front at some distance from the shoreline. The hybrid energy flux model's output is compared to that of another propagation model for this environment that is built on normal modes alone. General agreement in the shape, location, and amplitude of caustic features is observed with some discrepancies that may be attributable to inherent differences in the model derivations.  This work serves as a stepping-stone toward developing a more generalized three-dimensional energy flux model.
  \end{abstract}

  \section{Introduction}
  The classic two-dimensional (2D) energy flux model is a direct source-to-receiver calculation that avoids the need for finding eigenvalues, but it averages out the modal interference structure as originally derived \cite{weston1959guided}. One way of deriving the model involves inserting the locally-averaged WKB mode envelopes into a mode summation, expanding the incoherent intensity (or pressure-squared), and then transforming the summation over modenumber to an integration over propagation angle \cite{zhou2013integrating,harrison2013ray}.  This produces a locally-averaged depth-dependent intensity that tends to decay smoothly with range.  Range-dependence is handled by the adiabatic modes approximation and uses the ``ray invariant'', which is directly proportional to the closed phase integral, to map the propagation angles in range while the vertical transformation of propagation angles is handled by Snell's law.  This type of model when used for 3D environments is considered $N\times2D$ (N-by-2D) since it assumes azimuthal symmetry about the source location but is applied to several azimuthal angles independently, thus ignoring any horizontal refraction \cite{etter2018underwater}.

  Near-neighbor modal interference has been recently incorporated into the 2D energy flux model by the derivation of a convergence factor \cite{harrison2013ray}. The convergence factor resolves large-scale caustic features and shadow zones by focusing the contribution from a propagation angle when one of the ray families completes a cycle at the receiver location. This was an important development that bridged the gap between the classic incoherent energy flux model and the fully coherent normal mode solution.

  So far limited work has been done to develop three-dimensional (3D) energy flux models for ocean acoustic propagation \cite{weston1961horizontal,weston1980acoustic1,harrison1977three}, but such a model could potentially have significant computational advantages in complex environments or at high frequencies. This report derives a hybrid vertical normal mode and horizontal energy flux model used to solve for the 3D acoustic propagation in an ideal wedge environment with a radial sound speed front. The angular (vertical) modes are proper Sturm-Liouville (SL) eigenfunctions since we assume a pressure-release boundary condition at the sea surface and a rigid boundary condition at the seafloor. Thus the horizontal problem is completely separated and solved by the energy flux method for each angular mode.  Wentzel-Kramers-Brillouin (WKB) modes are used for the radial dimension and their summation is converted into an integration over propagation angle to construct the energy flux model for the horizontal problem.  A reflection coefficient for the radial sound-speed front is derived as well as the convergence factor.  The model is also extended to incorporate adiabatic range-dependence in the axial direction.

  Results from this hybrid energy flux model are compared to a fully normal mode model for the same environment \cite{lin2012analytical}. General agreement in the transmission loss (TL) is observed and the primary interference structure that propagates along the front is clearly observed. There are some mild discrepancies in overall TL and local features that may be attributable to the approximations used in the derivation of the model, but these are still being investigated for other sources of error. The purpose of deriving this three-dimensional energy flux model for the wedge environment is to demonstrate the generalizability of the energy flux approach and also serve as a stepping stone toward developing a more generalized three-dimensional energy flux model that can capture horizontal refraction in more realistic ocean environments.

  \section{Background}

  The energy flux method as originally derived for ocean acoustics can be understood as an incoherent mode sum that makes use of adiabatic modes to map the propagation angle distribution of intensity in a range-dependent stratified waveguide environment.  Weston in 1959 derived the ray invariant from considering a waveguide with slowly varying stratification structure and depth in range.  The derivation was obtained with three different approaches: one from ray theory, another from normal modes, and the final from energy conservation and reciprocity arguments \cite{weston1959guided}.  In two later papers published by Weston in 1980, a summary of energy flux concepts and formulas are discussed and used in a variety of analytical profile ducts as test cases \cite{weston1980acoustic1, weston1980acoustic2}.  Leonid Brekhovskikh in the Soviet Union was also investigating the incoherent intensity distribution in a cross-section of a depth-dependent stratified waveguide, and derived a range-independent energy flux model based on ray theory \cite{brekhovskikh1965average}.  Brekhovskikh's later book, \emph{Fundamentals of Ocean Acoustics} includes derivations of the ray invariant and occasional use of energy flux methods \cite{brekhovskikh2003fundamentals}.

  Other authors have also made contributions to the energy flux model.  Michael Milder in 1969 published a paper connecting Weston's ray invariant to action invariants derived with the stationary-action principle of classical and quantum mechanics \cite{milder1969ray}.  In 1973, P. W. Smith Jr. derived the energy flux model in terms of cycle distances for a slowly range-varying waveguide based on the theory of ray acoustics with lossy specular reflections from the boundaries \cite{smith1974averaged}.  Ji-Xun Zhou derived a closely related \emph{angular power spectrum} model from an incoherent mode sum of averaged depth-dependent mode envelopes for use in seabed scattering models and shallow-water long-range reverberation calculations \cite{zhou1980analytical, zhou2013integrating}.  Additionally, Charles Holland in 2010 used an energy flux model to derive an effective reflection loss for range-dependent incoherent intensity propagation \cite{holland2010propagation}. This effective reflection loss makes use of the geometric mean of the seabed plane-wave reflection coefficient and arithmetic mean of the cycle distance.

  Chris Harrison in the 1970's investigated use of the ray invariant for deriving analytical solutions of horizontal projections of ray paths in idealized 3D ocean acoustic environments and for the prediction of shadow zones in the horizontal plane \cite{harrison1977three, harrison1979acoustic}.  Of particular relevance to this paper, Harrison recently published two papers detailing the derivation of a \emph{convergence factor}, which reintroduces some near-neighbor modal interference (coherence) into the incoherent range-dependent energy flux model as a multiplicative factor inside of the angular integrand \cite{harrison2013ray, harrison2015efficient}.  This additional convergence factor is zero unless one or more of the four ray family cycles with source/receiver cycle offsets arrives at the receiver position. When a ray family completes a cycle at a receiver position, the convergence factor becomes non-zero and modulates caustic-like features which are usually present in high-frequency acoustic propagation models but are absent in the classic incoherent energy flux model.

  To investigate the potential use of energy flux models in 3D ocean acoustic propagation, canonical benchmark environments where 3D effects and horizontal refraction can occur were considered potential candidates for model derivation.  These environmental scenarios include propagation within a shoreline wedge, around conical seamounts, and across a V-shaped trough.  The wedge environment with a radial sound speed front as utilized in Lin and Lynch 2012 \cite{lin2012analytical} seemed an appropriate problem within which the energy flux method could be analogously implemented.  Lin and Lynch derived a solution from a normal mode expansion that made use of the endpoint method to construct a Green's function for the separated radial problem.  The smoothness boundary conditions at the frontal interface are satisfied and then the eigenvalues (wavenumbers) corresponding to discrete modes are found in the complex wavenumber-plane.

  The approach presented in this energy flux paper begins with an eigenbasis expansion of the angular (vertical) modes and then solves for the horizontally dependent modal coefficients using a semi-coherent energy flux approach.  Therefore this 3D propagation model still uses a 2D energy flux method but doesn't have the assumption of a 3rd collapsed dimension due to azimuthal symmetry. In addition, this approach for this problem requires using WKB approximations for a Bessel-like equation representing cylindrical spreading of waves instead of the usual depth-dependent WKB modes of most $N\times2D$ models.

  \section{Model Derivation}

\begin{figure}[ht]
    \includegraphics[width=\columnwidth]{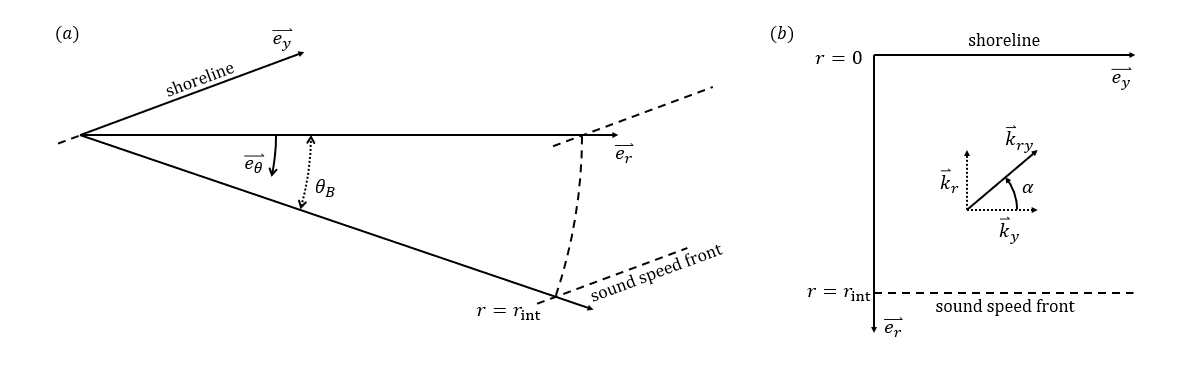}
    \caption{\label{fig:wedgeGeometryCartoon}{(a) Isometric view of wedge coordinates (b) Top-down view of r-y plane with propagation angle $\alpha$ defined. Adapted from Lin and Lynch 2012 \cite{lin2012analytical}.}}
  \end{figure}

  The wedge environment coordinate system is depicted in Fig.[\ref{fig:wedgeGeometryCartoon}] \cite{lin2012analytical}.  The bottom of the waveguide (at $\theta=\theta_B$) is assumed rigid and perfectly reflecting, a homogeneous Neumann boundary condition, while the surface is treated as pressure-release, a homogeneous Dirichlet boundary condition.  The frontal interface (at $r=r_{\mathrm{int}}$) divides the wedge environment into two regions: an isospeed duct ($c_1$ for $r<r_{\mathrm{int}}$) and another isospeed region radiating towards the $+r$-direction ($c_2$ for $r>r_{\mathrm{int}}$).  Thus the radial domain is from $r=0$ to $r=r_{\mathrm{int}}$, with a finiteness boundary condition at $r=0$ (the solution must be finite), and a homogeneous Robin boundary condition at $r=r_{\mathrm{int}}$.  Waves are also free to propagate in both the positive and negative $y$-directions, thus the boundary conditions for the $y$-domain are the Sommerfeld radiation conditions at $y=\pm\inf$. For use in derivations of the horizontal coefficient, we define a propagation angle, $\alpha$, embedded in the $r$-$y$ plane and measured from the $y$-parallel such that $k_y=k_{ry}\cos(\alpha)$.

  We will be solving for the \emph{transmission loss} (TL) which is defined as the logarithm of the ratio of intensity to a reference intensity.

  \begin{equation}\begin{split}
      \mathrm{TL} &= -10\log_{10}\(\frac{\mathcal{Z}_0}{\mathcal{Z}}\frac{\mathscr{P}}{\mathscr{P}_{\mathrm{ref}}}\) \quad\text{where}\quad \mathscr{P} = \abs{p}^2 \\
  \end{split}\end{equation}

  The governing wave equation can be written generally regardless of environment and coordinate geometry.  We assume time-harmonic solutions and use a time Fourier transform to obtain the 3D Helmholtz equation.

  \begin{equation}\begin{split}
      \[\lap - \frac{1}{c^2}\dnd{2}{}{t}\] p(\bvec{x},t) &= S(t)\delta\(\bvec{x}-\bvec{x_0}\) \\
      \Raro\qquad \[\lap + k^2\]p(\bvec{x},k) &= S(k)\delta\(\bvec{x}-\bvec{x_0}\) \\
  \end{split}\end{equation}

  The free-field solution of this inhomogeneous partial differential equation is simply the free-field Green's function.  This solution evaluated at $\unit[1]{m}$ distance will define the reference intensity.  If we set the monopole amplitude such that the free-field solution is unity at unit distance, then the reference pressure squared will also be unity and pressure-squared is then called $\mathscr{P}_{\mathrm{TL}}$ \cite{jensen2011computational, pierce2019acoustics}.

  \begin{equation}\begin{split}
      \[\lap + k^2\]G(\bvec{x}|\bvec{x_0}) &= \delta(\bvec{x}-\bvec{x_0}) \\
      \Let R &= \abs{\bvec{x}-\bvec{x_0}} \\
      G(\bvec{x}|\bvec{x_0}) &= \frac{e^{ikR}}{-4\pi R} \\
      \absevalat{S \cdot G(\bvec{x}|\bvec{x_0})}{R=1} &= S/4\pi \\
      \absevalat{S \cdot G(\bvec{x}|\bvec{x_0})}{R=1} &= 1 \quad\raro\quad S_{TL} = 4\pi \\
      \Raro\quad \mathrm{TL} &= -10\log_{10}\(\mathscr{P}_{\mathrm{TL}}\) \quad\tif\quad \mathcal{Z}=\mathcal{Z}_0 \\
  \end{split}\end{equation}

  Thus we seek to find the incoherent pressure-squared as scaled by a monopole amplitude of $4\pi$, ten times the logarithm of which is equivalent to the transmission loss.  To do so we begin by adopting the lateral cylindrical coordinate system and defining the boundary conditions.

  \begin{equation}\begin{split}
      \Let p(\bvec{x}) = p(r,\theta,y) &\tand \bvec{x_0} = (r_0,\theta_0,0) \\
      \lap p(\bvec{x}) + k^2 p(\bvec{x}) &= S\delta(\bvec{x}-\bvec{x_0}) \\
      k = \omega/c(r) &\tand c(r) = \left\{\begin{array}{ll} c_1 & \quad r<r_I \\ c_2 & \quad r\ge r_I\end{array}\right. \\
      \BCs p(r,0,y) = 0 &\tand \left.\dd{p}{\theta}\right|_{\theta=\theta_B} = 0 \\
  \end{split}\end{equation}

  Since the boundary conditions only depend on $\theta$, the homogeneous solution is assumed to be separable \cite{frisk1994ocean, lin2012analytical}.

  \begin{equation}\begin{split}
      \delta(\bvec{x}-\bvec{x_0}) &\stackrel{\mathrm{cyl}}{=} \frac{\delta(r-r_0)}{r}\delta(\theta-\theta_0)\delta(y) \\
      \lap_{\mathrm{cyl}} &= \frac{1}{r}\pp{}{r}\(r\pp{}{r}\) + \frac{1}{r^2}\pnp{2}{}{\theta} + \pnp{2}{}{y} \\
      r^{-1}\p_r\(r\p_rp\) &+ r^{-2}\p_\theta^2p + \p_y^2p + k^2p = S\frac{\delta(r-r_0)}{r}\delta(\theta-\theta_0)\delta(y) \\
  \end{split}\end{equation}

  \begin{equation}\begin{split}
      \text{Homogeneous:}\quad r^{-1}\p_r\(r\p_rp\) + r^{-2}\p_\theta^2p + \p_y^2p + k^2p &= 0 \\
      \text{Ansatz:}\qquad p = SA(r,y)\Theta(\theta) \\
      r^{-1}\p_r\(r\p_r(A\Theta)\) + r^{-2}\p_\theta^2(A\Theta) + \p_y^2(A\Theta) + k^2(A\Theta) &= 0 \\
      A^{-1}r\p_r\(r\p_r(A)\) + \Theta^{-1}\p_\theta^2(\Theta) + A^{-1}r^2\p_y^2(A) + k^2r^2 &= 0 \\
  \end{split}\end{equation}

  We see that some terms vary only with $\theta$ and other terms vary only with $r$ or $y$.  The only function that can be equivalent as $\theta$, $r$, and $y$ vary independently is a constant, thus we introduce the separation constant $\vkap^2$.

  \begin{equation}\begin{split}
      \Raro\qquad \Theta^{-1}\p_\theta^2(\Theta) = -\vkap^2 \\
      \p_\theta^2{\Theta} + \vkap^2\Theta = 0 \\
      \Theta = \CC_1\sin(\vkap\theta) + \CC_2\cos(\vkap\theta) \\
      \Theta(0)=0 \quad\Raro\quad \Theta = \CC_1\sin(\vkap\theta) \\
      \evalat{\p_\theta\Theta}{\theta=\theta_B} = 0 \quad\Raro\quad \vkap_n = \frac{(n-\half)\pi}{\theta_B} \\
  \end{split}\end{equation}

  In fact, these homogeneous boundary conditions satisfy a regular Sturm-Liouville problem, so we begin by constructing a normalized eigenbasis expansion in the $\theta$-coordinate, i.e. we find homogeneous solutions satisfying the boundary conditions \cite{arfken2012mathematical, haberman1983elementary}.

  \begin{equation}\begin{split}
      \Theta_n(\theta) &= \CC_n\sin\(\vkap_n\theta\) \\
      \int_0^{\theta_B} \Theta_n\Theta_m\d\theta = \delta_{mn} \quad&\Raro\quad \int_0^{\theta_B} \CC_n^2\sin^2(\vkap_n\theta) \d\theta = 1 \\
      \CC_n^2\int_0^{\theta_B}(1-\cos(2\vkap_n\theta))\d\theta = 2 \quad&\Raro\quad \CC_n^2\[\theta-(2\vkap_n)^{-1}\sin(2\vkap_n\theta)\]_0^{\theta_B} = 2 \\
      \CC_n^2\[\theta_B - (2\vkap_n)^{-1}\sin(2\vkap_n\theta_B)\] = 2 \quad&\Raro\quad \CC_n^2\[\theta_B - \cancel{(2\vkap_n)^{-1}\sin((2n-1)\pi)}\] = 2 \\
      \CC_n &= \sqrt{2/\theta_B} \\
      \Theta_n(\theta) = \sqrt{2/\theta_B}\sin(\vkap_n\theta) &\twhere \vkap_n = (n-\half)\pi/\theta_B \\
  \end{split}\end{equation}

  Now we can expand the solution of the inhomogeneous 3D helmholtz equation in a normalized eigenbasis of $\theta$.  Generally speaking, for a regular Sturm-Liouville problem, the differential operator is self-adjoint and results in the eigenvalue when applied to an eigenfunction \cite{arfken2012mathematical, haberman1983elementary}.

  \begin{equation}\begin{split}
      \braket{f(x)}{g(x)} &\equiv \int_{-\inf}^{+\inf}f^*(x)g(x)\dx \\
      \text{generally:}\quad f(\theta) &= \sum_m\braket{\Theta_m}{f}\Theta_m(\theta) \\
      \delta(\theta-\theta_0) &= \sum_m\braket{\Theta_m}{\delta(\theta-\theta_0)}\Theta_m(\theta) \\
      &= \sum_m\Theta_m^*(\theta_0)\Theta_m(\theta) \\
      \Let \mathscr{L}_\theta &= r^{-2}\pnp{2}{}{\theta} \\
      \Let \mathscr{L}_{ry} &= \frac{1}{r}\pp{}{r}\(r\pp{}{r}\) + \pnp{2}{}{y} \\
  \end{split}\end{equation}

  \begin{equation}\begin{split}
      \Ansatz p = \sum_mSA_m(r,y|r_0,\theta_0)&\Theta_m(\theta) \\
      \[\mathscr{L}_{ry} + \mathscr{L}_\theta + k^2\]p &= S\frac{\delta(r-r_0)}{r}\delta(\theta-\theta_0)\delta(y) \\
      \[\mathscr{L}_{ry} + \mathscr{L}_\theta + k^2\]\sum_mSA_m\Theta_m(\theta) &= S\frac{\delta(r-r_0)}{r}\delta(y)\sum_m\Theta_m^*(\theta_0)\Theta_m(\theta) \\
      S\sum_m\[\mathscr{L}_{ry} A_m - \frac{\vkap_m^2}{r^2} A_m + k^2 A_m\] \Theta_m(\theta) &= S\sum_m\frac{\delta(r-r_0)}{r}\delta(y)\Theta_m^*(\theta_0)\Theta_m(\theta) \\
      \Let A_m(r,y|r_0,\theta_0) &= \Theta_m(\theta_0)B_m(r,y|r_0) \\
      S\sum_m\[\mathscr{L}_{ry} B_m - \frac{\vkap_m^2}{r^2} B_m + k^2 B_m\] \Theta_m^*(\theta_0)\Theta_m(\theta) &= S\sum_m\[\frac{\delta(r-r_0)}{r}\delta(y)\]\Theta_m^*(\theta_0)\Theta_m(\theta) \\
  \end{split}\end{equation}

  $A_m$ must contain the $\Theta(\theta)$ normal mode evaluated at the source location $\theta_0$. We pull this factor out directly and note that for each modenumber $m$ we have a new differential equation to solve for the coefficient, now $B_m(r,y|r_0)=A_m(r,y|r_0,\theta_0)/\Theta_m(\theta_0)$. Note also that we will primarily be considering real valued modefunctions and therefore will drop the complex conjugation in the inner product. We will also temporarily drop the $m$ subscript, with the understanding that we must solve an analogous differential equation to obtain the horizontally-dependent modal coefficient for each modenumber $m$.

  \begin{equation}\begin{split}
      \[\mathscr{L}_{ry} + \(k^2-\frac{\vkap^2}{r^2}\)\]B(r,y|r_0) &= \frac{\delta(r-r_0)}{r}\delta(y) \\
      \[\frac{1}{r}\pp{}{r}\(r\pp{}{r}\) + \pnp{2}{}{y} + k_{ry}^2\]B(r,y|r_0) &= \frac{\delta(r-r_0)}{r}\delta(y) \\
  \end{split}\end{equation}

  Before moving on, we should note in advance that this horizontal problem is also separable.  Acoustic energy freely propagates in the positive and negative $y$-directions, and the remaining boundary conditions depend on either $y$ or $r$ alone.  By introducing a separation constant for each horizontal problem dependent on mode $m$ (the subscript $m$ we will temporarily omit for clarity), $-k_y^2$, we can immediately find two separable differential equations so that our solution is assumed to be constructed as a product of orthogonal solutions, $B(r,y|r_0) = R(r|r_0)Y(y|0)$.

  \begin{equation}\begin{split}
      \Let B(r,y|r_0) = R(r|r_0)Y(y|0) & \\
      \[\frac{1}{r}\pp{}{r}\(r\pp{}{r}\) + \pnp{2}{}{y} + k_{ry}^2\]RY &= 0 \\
      Y\frac{1}{r}\pp{}{r}\(r\pp{R}{r}\) + R\pnp{2}{Y}{y} + k_{ry}^2RY &= 0 \\
      R^{-1}\frac{1}{r}\pp{}{r}\(r\pp{R}{r}\) + Y^{-1}\pnp{2}{Y}{y} + k_{ry}^2 &= 0 \\
      Y^{-1}\pnp{2}{Y}{y} = -k_y^2 \\
      R^{-1}\frac{1}{r}\pp{}{r}\(r\pp{R}{r}\) + k_{ry}^2 = +k_y^2 \\
  \end{split}\end{equation}

  This horizontal problem we intend to solve using the energy flux method, which we will construct from a WKB mode summation. We assume that we have a duct extending from $r=0$ to $r=r_{\mathrm{int}}$, which is the mathematical domain upon which we define our solution space. At $r=0$ we require that the solution be finite, and at the interface there is a sound speed transition from $c_1$ to $c_2$. We will assume that there are normal modes (eigenfunctions) that can expand the solution into a spectral decomposition, which we shall see is justified since the radial problem is a singular Sturm-Liouville problem for which we can find an eigenbasis \cite{arfken2012mathematical, haberman1983elementary}.  The effective wavenumber in the horizontal problem is no longer invariant in $r$ (i.e. $k_{ry}^2=k^2-k_\theta^2=k^2-\nicefrac{\vkap^2}{r^2}$). Since we are presently assuming that $k$ is constant before the interface, this ODE is actually a scaled version of Bessel's differential equation. To allow for mild $r$ dependence of the bulk wavenumber $k(r)$, we will use WKB modes to approximate the theoretically exact radial modes. Once the mode sum is constructed, we will use derivatives of the \emph{closed} phase integral and a Snell's law analog (invariance of $k_y$) to map the mode summation to an integration over propagation angle \cite{harrison2010fixed, harrison2013ray}.

  \begin{equation}\begin{split}
      \Let k_r^2 &= k_{ry}^2 - k_y^2 = k^2 - \nicefrac{\vkap^2}{r^2} - k_y^2 \\
      \p_y^2Y + k_y^2Y &= 0 \\
      \p_r^2R + r^{-1}\p_rR + k_{r}^2R &= 0 \\
      B(r,y|y_0) &= \sum_n Y_n(y|0)R_n(r|r_0)\quad \text{where $Y_n$ is the modal coefficient} \\
  \end{split}\end{equation}

  The classic energy flux model can be derived from an incoherent mode sum in the semi-classical Wentzel-Kramers-Brillouin (WKB) approximation.  WKB modes is often interpreted as a locally-planar wave approximation which requires that the relative change in vertical wavenumber is small on the scale of the vertical wavelength \cite{brekhovskikh2003fundamentals, jensen2011computational}.  This approach assumes the solution as an exponential function with amplitude and phase functions constructed of perturbation series.  Terms of similar perturbation order are equated in the differential equation expansion, leading to a series of high-frequency acoustic approximations \cite{bender2013advanced}.  In order to use this formulation, we look for a small parameter $\veps$ to expand by. This small parameter will likely be related to $1/k = \lambda/2\pi$, which we will simply call $\lambda_0$ for the present moment.

  \newcommand{\lamo}{\lambda_0}

  \begin{equation}\begin{split}
      \Let \lamo &= 1/k = \lambda/2\pi \tand k_r = \eta_r/\lamo\\
      \lamo^2 R'' + \lamo^2 r^{-1}R' &= -\eta_r^2R \\
      \Let R(r) &\sim \exp\left[\frac{1}{\veps}\sum_{n=0}^{\infty}\veps^nS_n(r)\right] \\
      R' &= \(\frac{1}{\veps}\sum_{n=0}^\inf\veps^nS_n'\)R \\
      R'' &= \(\frac{1}{\veps}\sum_{n=0}^\inf\veps^nS'_n\)^2R + \(\frac{1}{\veps}\sum_{n=0}^\inf\veps^nS''_n\)R \\
  \end{split}\end{equation}

  Plugging the series expanded derivatives into the perturbed differential equation, we note that the coefficients of the exponentials must be equal to zero.  Note that in the perturbed differential equation the parameter is multiplied to the second derivative, so we must be finding an approximation such that the curvature of this function is negligible.  We will be evaluating the "dominant balance" in terms of the order of the small parameter $\lamo$.

  \begin{equation}\begin{split}
      \lamo^2 R'' + \lamo^2 r^{-1}R' = -\eta_r^2R & \\
      \left\{\lamo^2\(\frac{1}{\veps}\sum_{n=0}^\inf\veps^nS'_n\)^2 + \veps\(\frac{1}{\veps}\sum_{n=0}^\inf\veps^nS''_n\) + r^{-1}\(\frac{1}{\veps}\sum_{n=0}^\inf\veps^nS_n'\)\right\}R &= -\eta_r^2R \\
      \frac{\lamo^2}{\veps^2}\(\sum_{n=0}^\inf\veps^nS'_n\)^2 + \frac{\lamo^2}{\veps}\(\sum_{n=0}^\inf\veps^nS''_n\) + \frac{\lamo^2}{\veps}r^{-1}\(\sum_{n=0}^\inf\veps^nS_n'\) &= -\eta_r^2 \\
  \end{split}\end{equation}

  We assume that the order of $-\eta_r^2$ and the lowest $\veps$-order term are both of unity order, $\mathscr{O}(1)$.  This implies that $\lambda_0$ is likely $\mathscr{O}(\veps)$ in our construction.

  \begin{equation}\begin{split}
      \frac{\lamo^2}{\veps^2}((S_0')^2+2\veps S_0'S_1'+...) + \frac{\lamo^2}{\veps}(S_0''+\veps S_1''+...) + \frac{\lamo^2}{\veps}r^{-1}(S_0'+\veps S_1'+...) &= -\eta_r^2 \\
      \(\frac{\lamo^2}{\veps^2}(S_0')^2+\frac{\lamo^2}{\veps}2S_0'S_1'+...\) + \(\frac{\lamo^2}{\veps}S_0''+\lamo^2S_1''+...\) + r^{-1}\(\frac{\lamo^2}{\veps}S_0'+\lamo^2S_1'+...\) &= -\eta_r^2 \\
      \Let \lamo\sim\veps \tand \frac{\lamo^2}{\veps^2}\sim-\eta_r^2 \\
      \lamo^0(S_0')^2 + \lamo^1\(2S_0'S_1'+S_0''+r^{-1}S_0'\) + ... &= -\eta_r^2 \\
  \end{split}\end{equation}

  We assume that the left and right hand sides of this equation must be equivalent for all scales, i.e. $\mathscr{O}(\lamo^n)$.  Thus by dominant order we obtain a series of differential equations which allow us to solve for the phase function terms $S_n$ \cite{bender2013advanced}.

  \begin{equation}\begin{split}
      (S_0')^2 &= -\eta_r^2 \\
      S_0'' + 2S_0'S_1' + r^{-1}S_0' &= 0 \\
      S_0 &= \int\d S_0 = \int S_0'\dr \\
      &= \pm i\int\eta_r\dr \\
      S_0' &= \pm i\eta_r \\
      S_0'' &= \pm i\eta_r' \\
  \end{split}\end{equation}

  \begin{equation}\begin{split}
      \pm i\eta_r' \pm 2i\eta_rS_1' \pm ir^{-1}\eta_r &= 0 \\
      \eta_r' + 2\eta_r S_1' + r^{-1}\eta_r &= 0 \\
      S_1' &= -\half\[\frac{\eta_r'}{\eta_r}+\frac{1}{r}\] \\
      S_1 &= \int\d S_1 = \int S_1'\dr \\
      &= -\half\int\[\frac{\eta_r'}{\eta_r}+\frac{1}{r}\]\dr \\
      &= -\half\[\int\eta_r^-1\d \eta_r + \int r^{-1}\dr\] \\
      \NB &\eta_r>0 \tand r>0 \\
      S_1 &= -\half\[\ln(\eta_r)+\ln(r)+\CC_0\] \\
  \end{split}\end{equation}

  Now with these phase function terms, we substitute back into the asymptotic expansion of R.  Note that the exponential function contains the \emph{open} phase integral which is over the dummy variable $\rho$, and is evaluated from an arbitrary reference point $r_{\mathrm{ref}}$. It is useful to note here that since we don't know the actual form of $k_r(r)$, we represent the indefinite integral of it including an arbitrary integration constant by writing it as a definite integral from an arbitrary reference point to the dependent variable.

  \begin{equation}\begin{split}
      R(r) &\sim \exp\[\pm\frac{i}{\lamo}\int_{r_{\mathrm{ref}}}^r\eta_r(\rho)\d\rho-\half(\ln(\eta_r)+\ln(r)+\CC_0)\] \twhile \lamo\raro0^+ \\
      &\sim \exp\[\pm\frac{i}{\lamo}\int_{r_{\mathrm{ref}}}^r\eta_r(\rho)\d\rho + \ln\(\eta_r^{-\half}\) + \ln\(r^{-\half}\) + \CC_1\] \\
      &\sim \[\eta_r r\]^{-\half}\CC_2\exp\[\pm i\lamo^{-1}\int_{r_{\mathrm{ref}}}^r\eta_r(\rho)\d\rho\] \\
      R &\sim \CC_3\[\eta_r r\]^{-\half}\exp\[+i\int_{r_{\mathrm{ref}}}^r k_r(\rho)\d\rho\] + \CC_4\[\eta_r r\]^{-\half}\exp\[-i\int_{r_{\mathrm{ref}}}^r k_r(\rho)\d\rho\] \\
  \end{split}\end{equation}

  Since we are working with a singular Sturm-Liouville problem, we shall choose to normalize our eigenfunctions in the $\mathcal{L}_2$-norm. The domain of this problem we define as $r\in[0,r_{\mathrm{int}}]$, and the boundary conditions are finiteness at $r=0$ and a Robin boundary condition at $r=r_{\mathrm{int}}$. These boundary conditions will satisfy the requirements that our differential operator is self-adjoint, i.e. that the boundary terms vanish when applying integration by parts to the ODE over its domain. Going back to our radial differential equation, we note that matching the Sturm-Liouville form highlights the presence of a weighting function that must be accounted for in the normalization \cite{arfken2012mathematical, haberman1983elementary}.

  \begin{equation}\begin{split}
      \text{SL form:}\qquad \p_x\[p(x)\p_xf(x)\]+q(x)f(x)+\lambda w(x)f(x) &= 0 \\
      \text{BDE form:}\qquad r^{-1}\p_r\[r\p_rR\] + k_ry^2R + k_y^2R &= 0\\
      \p_r\[r\p_rR\] + k_{ry}^2rR + k_y^2rR &= 0 \\
      \p_r\[p(r)\p_rR\]+q(r)R+\lambda w(r)R &= 0 \\
      p(r)=r \qquad q(r)=k_{ry}^2 \qquad w(r) &= r \\
      \text{SL normalization:}\qquad \int_a^b w(r)R^*(r)R(r)\dr = 1 \\
  \end{split}\end{equation}

  \newcommand{\Ec}{\mathcal{E}}
  \newcommand{\Ecp}{\mathcal{E}_+}
  \newcommand{\Ecm}{\mathcal{E}_-}

  \begin{equation}\begin{split}
      \Let \phi = \int_{r_{\mathrm{ref}}}^r k_r(\rho)\d\rho \\
      \int_a^b r\[\eta_rr\]^{-1}\[(\CC_3e^{+i\phi}+\CC_4e^{-i\phi})(\CC_3^*e^{-i \phi}+\CC_4^*e^{+i\phi})\]\dr &= 1 \\
      \int_a^b \eta_r \[\CC_3\CC_3^*+\CC_3\CC_4^*e^{+i2\phi}+\CC_4\CC_3^*e^{-i2\phi}+\CC_4\CC_4^*\]\dr &= 1 \\
      \int_a^b \eta_r \[\abs{\CC_3}^2+\abs{\CC_4}^2+\CC_3\CC_4^*\(\cos(2\phi)+i\sin(2\phi)\)+\CC_4\CC_3^*\(\cos(2\phi)-i\sin(2\phi)\)\]\dr &= 1 \\
  \end{split}\end{equation}

  \begin{equation}\begin{split}
      \Let \Gamma = \int_{r_{\mathrm{ref}}}^rk_r(\rho)\d\rho &\tand \Phi = \Delta\Gamma \\
      \Ansatz \CC_3e^{+i\Gamma} + \CC_4e^{-i\Gamma} &= \CC_5\cos\(\Gamma+\Phi\) \\
      &= \half \CC_5 \(e^{+i\Gamma}e^{+i\Phi}+e^{-i\Gamma}e^{-i\Phi}\) \\
      \Let \CC_3=\CC_3^\rho\exp\[i\CC_3^\phi\] \\
      \CC_3^\rho\exp\[+i\Gamma+i\CC_3^\phi\] + \CC_4^\rho\exp\[-i\Gamma+i\CC_4^\phi\] &= \half\CC_5^\rho\exp\[+i\Gamma+i\Phi+i\CC_5^\phi\] + \half\CC_5^\rho\exp\[-i\Gamma-i\Phi+i\CC_5^\phi\] \\
      \CC_3^\rho = \half\CC_5^\rho &\tand \CC_4^\rho = \half\CC_5^\rho \\
      \CC_3^\phi = \CC_5^\phi + \Phi &\tand \CC_4^\phi = \CC_5^\phi - \Phi \\
      \Raro\qquad \CC_3^\rho = \CC_4^\rho &\tand \CC_5^\rho = 2\CC_3^\rho = 2\CC_4^\rho \\
      \Raro\qquad \CC_5^\phi = \half(\CC_3^\phi+\CC_4^\phi) &\tand \Phi = \half(\CC_3^\phi-\CC_4^\phi) \\
  \end{split}\end{equation}

  By making the normally reasonable assumption that the refracted wave returns with the same amplitude, we have shown that it would be arbitrarily possible to express this sum of complex exponential oscillations as a cosine function with phase shift $\Phi$.

  \begin{equation}\begin{split}
      \Let R(r) \sim \CC_5\[\eta_r r\]^{-\half}\cos\(\Gamma+\Phi\)& \\
      \text{SL normalization:}\qquad \int_a^b w(r)R^*(r)R(r)\dr = 1 \\
      \int_a^b r\frac{\abs{\CC_5}^2}{\eta_r r}\cos^2\(\Gamma+\Phi\)\dr &= 1 \\
      \text{Assume $\cos^2()$ has an average value of 1/2} \\
      \int_a^b \frac{\abs{\CC_5}^2}{\eta_r}\dr &= 2 \\
  \end{split}\end{equation}

  \begin{equation}\begin{split}
      \Let \mathfrak{L}_\eta &= \int_a^b\frac{1}{\eta_r}\dr \\
      \abs{\CC_5} &= \sqrt{\frac{2}{\mathfrak{L}_\eta}} \\
      R(r) &\sim \sqrt{\frac{2}{\mathfrak{L}_\eta}}\[\eta_r r\]^{-\half}\cos\[\int_{r_{\mathrm{ref}}}^rk_r(\rho)\d\rho + \Phi\] \\
  \end{split}\end{equation}

  Note that we can reinsert the quantity $\lamo$ since the normalization will balance out the scale and we still have a normalized mode function.

  \begin{equation}\begin{split}
      k_r^2 &= \eta_r^2/\lamo^2 \\
      \Let \mathfrak{L}_k &= \int_a^b\frac{1}{k_r}\dr \\
      R(r) &\sim \sqrt{\frac{2}{\mathfrak{L}_\eta}}\sqrt{\frac{\lamo}{\lamo}}\[\eta_r r\]^{-\half}\cos\[\int_{r_{\mathrm{ref}}}^rk_r(\rho)\d\rho + \Phi\] \\
      R(r) &\sim \sqrt{\frac{2}{\mathfrak{L}_k}}\[k_r r\]^{-\half}\cos\[\int_{r_{\mathrm{ref}}}^rk_r(\rho)\d\rho + \Phi\] \\
  \end{split}\end{equation}

  The cycle distance for a ray is related to this integral value, $\mathfrak{L}_k$, for which we will drop the subscript $k$ designation now. The differential line element for cylindrical coordinates has no Jacobian scaling terms on the differential $\dr$ and differential $\dy$ elements. Thus the path integral essentially evaluates the same way as in Cartesian coordinates \cite{arfken2012mathematical}. Recall from Fig.[\ref{fig:wedgeGeometryCartoon}] that $\alpha$ is the propagation angle in the $r,y$-plane measured from the parallel to the shoreline (along the wedge apex).

  \begin{equation}\begin{split}
      \ds^2 &= \dr^2 + r^2\d\theta^2 + \dy^2 \\
      \DD &= 2\int_{r'}^{r''}\dy \\
      &= 2\int_{r'}^{r''}\dd{y}{r}\dr \\
      &= 2\int_{r'}^{r''}\cot(\alpha)\dr \\
      &= 2\int_{r'}^{r''}\frac{k_y}{k_r}\dr \\
      &= 2k_y\mathfrak{L} \\
      \mathfrak{L} &= \frac{\DD}{2k_y} \\
  \end{split}\end{equation}

  Thus we may re-express the WKB mode function in terms of the cycle distance, which is simply another cycle integrated quantity that is directly related to the mode normalization \cite{brekhovskikh2003fundamentals}.

  \begin{equation}\begin{split}
      R(r) &\sim 2\sqrt{\frac{k_y}{\DD}}\[k_r r\]^{-\half}\cos\[\int_{r_{\mathrm{ref}}}^rk_r(\rho)\d\rho + \Phi\] \\
  \end{split}\end{equation}

  We have obtained our approximate modefunctions, but we have not calculated any eigenvalues.  Finding eigenvalues typically involves applying boundary conditions to this function (either as reflection or refraction) on both sides of the domain, thus constructing a \emph{closed} phase integral where allowable values of $k_y$ may be obtained by satisfying this characteristic equation \cite{bender2013advanced, brekhovskikh2003fundamentals}.  In our case it is enough to assume that there are eigenvalues $k_y$ which are directly tied to the definition of this WKB modefunction, i.e. $k_y$ embedded in the definition of $k_r$.

  For a singular Sturm-Liouville problem (like Bessel's Differential Equation), we can construct an eigenbasis expansion of the Dirac delta distribution which we may consider to be a best approximation within our eigenspace.  By using the definition of eigenfunction expansion with the orthogonality condition that includes the weighting function for the Sturm-Liouville problem ($w(r)=r$) we find the closure relation for the Dirac delta function within our basis \cite{arfken2012mathematical}.

  \begin{equation}\begin{split}
      w(r) &= r \\
      \frac{\delta(r-r_0)}{r} &\sim \sum_{n=1}^\inf a_n R_n(r) \\
      \int_{0}^{r_{\mathrm{int}}}\frac{\delta(r-r_0)}{r}R_m(r)w(r)\dr &= \int_{0}^{r_{\mathrm{int}}}\sum_{n=1}^\inf a_nR_n(r)R_m(r)w(r)\dr \\
      \text{Evaluates to 0 unless }n &= m \\
      \int_{0}^{r_{\mathrm{int}}}\delta(r-r_0)R_n(r)\dr &= a_n \int_{0}^{r_{\mathrm{int}}}rR_n^2(r)\dr \\
      a_n &= \frac{\int_{0}^{r_{\mathrm{int}}}\delta(r-r_0)R_n(r)r\dr}{\int_{0}^{r_{\mathrm{int}}}rR_n^2(r)\dr} \\
      \text{eigenfunctions are normalized, }\int_0^{r_{\mathrm{int}}} rR_n^2\dr &= 1 \\
      a_n &= \int_{0}^{r_{\mathrm{int}}}\delta(r-r_0)R_n(r)\dr \\
      a_n &= R_n(r_0) \\
      \Raro\qquad \frac{\delta(r-r_0)}{r} &\sim \sum_{n=1}^\inf R_n(r_0)R_n(r) \\
  \end{split}\end{equation}

  Since we have expanded the horizontal problem as an eigenbasis expansion of radial WKB modes, the last piece we must find is the modal coefficients dependent on the $y$-direction.  Since we must solve this problem for each radial modefunction, we assume that $k_y$ is selected at this point, thus we no longer have an eigenvalue problem, but a deterministic inhomogeneous ODE forced by an impulse term.  Thus it is the equation for the Green's function in the $y$-direction, solvable by the endpoint method.  The endpoint method can be thought of as the method of variation of parameters applied to an ODE with a Dirac delta forcing term, thus constructing a piecewise continuous solution that satisfies the boundary conditions on either side, continuity at the source location, and the first derivative jump condition at the source location \cite{frisk1994ocean, arfken2012mathematical}.

  \begin{equation}\begin{split}
      \Let B(r,y|r_0) = \sum_n Y_n(y|0)R_n(r|r_0) & \\
      \[\frac{1}{r}\pp{}{r}\(r\pp{}{r}\) + \pnp{2}{}{y} + k_{ry}^2\]B &= \frac{\delta(r-r_0)}{r}\delta(y) \\
      \[\frac{1}{r}\pp{}{r}\(r\pp{}{r}\) + \pnp{2}{}{y} + k_{ry}^2\]\sum_nY_n(y|0)R_n(r_0)R_n(r) &= \sum_nR_n(r_0)R_n(r)\delta(y) \\
      \NB r^{-1}\p_r\(r\p_rR(r)\) &= -k_r^2R(r) \\
      \sum_nR_n(r_0)\[\frac{1}{r}\p_r\(r\p_rR_nY_n\)+\p_y^2R_nY_n+k_{ry}^2R_nY_n\] &= \sum_nR_n(r_0)\delta(y)R_n(r) \\
      \sum_nR_n(r_0)\[-k_r^2R_nY_n+R_n\p_y^2Y_n+k_{ry}^2R_nY_n\] &= \sum_nR_n(r_0)\delta(y)R_n(r) \\
      \sum_nR_n(r_0)\[\p_y^2Y_n+k_{y}^2Y_n\]R_n &= \sum_nR_n(r_0)\delta(y)R_n(r) \\
      \p_y^2Y_n+k_y^2Y_n &= \delta(y) \\
  \end{split}\end{equation}

  Now that we have expanded the solution in the $r$-dependent eigenbasis and separated the inhomogeneous differential equation into its $y$-component, we can proceed with obtaining the $y$-dependent Green's function.  The only boundary conditions we have in the $y$-direction problem are radiation conditions at $\pm\inf$, thus we choose a homogeneous solution (as complex exponentials) for either side of the domain (fixing the source at $y_0=0$) such that the waves are propagating in the appropriate direction to satisfy the Sommerfeld radiation condition \cite{arfken2012mathematical}.

  \begin{equation}\begin{split}
      \text{Inhomogeneous problem:}\qquad \[\p_y^2 + k_y^2\]Y(y|0) &= \delta(y) \\
      \text{Homogeneous problem:}\qquad \pnp{2}{Y}{y}+k_y^2Y &= 0 \\
      \text{Homogeneous solution:}\qquad Y &= Ae^{+ik_yy} + Be^{-ik_yy} \\
      \Let \pnp{2}{Y(y|y_0)}{y} + k_y^2Y(y|y_0) &= \delta(y-y_0) \\
  \end{split}\end{equation}

  \begin{equation}\begin{split}
      Y &= \left\{\begin{matrix}Ae^{-ik_y(y-y_0)};&\quad y<y_0 \\ Ae^{+ik_y(y-y_0)}; &\quad y>y_0\end{matrix}\right\} = Ae^{+ik_y|y-y_0|} \\
      \dd{Y}{y} &= \left\{\begin{matrix}-ik_y Ae^{-ik_y(y-y_0)};&\quad y<y_0 \\ +ik_y Ae^{+ik_y(y-y_0)}; &\quad y>y_0\end{matrix}\right\} = i\cdot\sgn\{y-y_0\}\cdot k_y Ae^{+ik_y|y-y_0|} \\
  \end{split}\end{equation}

  \begin{equation}\begin{split}
      p(y)=1 \tand \lim{\eps\raro0+}\[\left.\dd{Y}{y}\right|_{y=y_0+\eps}-\left.\dd{Y}{y}\right|_{y=y_0-\eps}\] &= \frac{1}{p(y_0)} \\
      \lim{\eps\raro0+}\[\left.\sgn\{y-y_0\}ik_y Ae^{+ik_y|y-y_0|}\right|_{y=y_0+\eps} - \left.\sgn\{y-y_0\}ik_y Ae^{+ik_y|y-y_0|}\right|_{y=y_0-\eps}\] &= 1 \\
  \end{split}\end{equation}

  \begin{equation}\begin{split}
      \lim{\eps\raro0+}\[ik_y Ae^{+ik_y\eps} + ik_y Ae^{+ik_y\eps}\] &= 1 \\
      i2k_y A &= 1 \\
      \Raro\qquad A &= \frac{-i}{2k_y} \\
      Y(y|0) &= \frac{-i}{2}\frac{e^{ik_y\abs{y}}}{k_y} \\
  \end{split}\end{equation}

  Thus with an eigenbasis for $\theta$ and $r$ directions and a Green's function for the $y$-direction, we plug these into the separable solution to obtain our pressure expressed as a mode sum.

  \begin{equation}\begin{split}
      \Theta_m(\theta) &= \sqrt{2/\theta_B}\cdot\sin(\vkap_n\theta) \twhere \vkap_n = (n-\half)\pi/\theta_B \\
      R_{mn}(r) &\sim 2\sqrt{\frac{k_{y|mn}}{\DD_{mn}}}\[k_{r|mn}r\]^{-\half}\cos\[\int_{r_{\mathrm{ref}}}^rk_{r|mn}(\rho)\d\rho + \Phi\] \\
      Y_{mn}(y) &= \frac{-i}{2}\frac{e^{ik_{y|mn}\abs{y}}}{k_{y|mn}} \\
      p &= S\sum_m \Theta_m(\theta_0)\Theta_m(\theta)B_m(r,y|r_0) \\
      &= S\sum_m\Theta_m(\theta_0)\Theta_m(\theta)\left\{\sum_nY_{mn}(y|0)R_{mn}(r_0)R_{mn}(r)\right\} \\
      &= \frac{-iS}{2}\sum_m\Theta_m(\theta_0)\Theta_m(\theta)\left\{\sum_n\frac{e^{ik_{y|mn}\abs{y}}}{k_{y|mn}}R_{mn}(r_0)R_{mn}(r)\right\} \\
  \end{split}\end{equation}

  The quantity $B_m(r,y|r_0)$ represents the horizontally dependent modal coefficient, which is the solution to the horizontal problem, and will be the energy flux solution once we turn the WKB mode summation into an integration over propagation angle. To do so we return to the incoherent mode sum that represents the pressure-squared. We must expand the modal cross product and specify which terms we neglect when calculating the incoherent pressure-squared.  For the angular(vertical) modes, we are presently concerned with only the incoherent product ($m=p$) so we neglect all cross-products.

  \begin{equation}\begin{split}
      \mathscr{P}_{\mathrm{TL}} &= \abs{p}^2 \\
      &= \abs{S\sum_m\Theta_m(\theta|\theta_0) B_m}^2 \\
      &= \[S\sum_m\Theta_m(\theta|\theta_0) B_m\]\times\[S^*\sum_p\Theta_p^*(\theta|\theta_0) B_p^*\] \\
      &= S^2\sum_m\abs{\Theta_m(\theta|\theta_0)}^2\abs{B_m}^2 \\
  \end{split}\end{equation}

  Again we have a mode summation within $B_m$, but we will at first discard the coherent cross-products and then reincorporate some of them later in the derivation of the convergence factor \cite{harrison2013ray}.

  \begin{equation}\begin{split}
      \abs{B_m}^2 &= \abs{\sum_n\frac{-ie^{ik_y\abs{y}}}{2k_y}R_{mn}(r|r_0)}^2 \\
      &= \[\sum_n\frac{-ie^{ik_{y|mn}\abs{y}}}{2k_{y|mn}}R_{mn}(r|r_0)\]\[\sum_q\frac{+ie^{-ik_{y|mq}\abs{y}}}{2k_{y|mq}}R_{mq}^*(r|r_0)\] \\
      &= \frac{1}{4}\(\sum_{n=1}^\inf\frac{\abs{R_{mn}(r|r_0)}^2}{k_{y|mn}^2} + \sum_{n=2}^\inf\sum_{q=1}^{n-1}\frac{R_{mn}(r|r_0)R_{mq}^*(r|r_0)}{k_{y|mn}k_{y|mq}} \right. \\
      & \qquad\qquad \left. \times \left\{\exp\[i(k_{y|mn}-k_{y|mq})\abs{y}\]+\exp\[-i(k_{y|mn}-k_{y|mq})\abs{y}\]\right\}\) \\
      &= \frac{1}{4}\(\sum_{n=1}^\inf\frac{\abs{R_{mn}(r|r_0)}^2}{k_{y|mn}^2} + \sum_{n=2}^\inf\sum_{q=1}^{n-1}\frac{R_{mn}(r|r_0)R_{mq}^*(r|r_0)}{k_{y|mn}k_{y|mq}} 2\cos\[(k_{y|mn}-k_{y|mq})\abs{y}\]\) \\
  \end{split}\end{equation}

  We now have a workable expression for the incoherent pressure-squared, obtained by disregarding all coherent cross-products. To convert this to energy flux, we will assume the use of \emph{locally-averaged mode envelopes} \cite{harrison2010fixed, zhou2013integrating}.  Essentially we are averaging out the contribution from the oscillating $\cos^2()$, replaced by a constant factor of $1/2$.

  \begin{equation}\begin{split}
      \mathscr{P}_{\mathrm{TL}} &= S^2\sum_m\abs{\Theta_m(\theta|\theta_0)}^2\left\{\frac{1}{4}\sum_{n=1}^{\inf}\frac{\abs{R_{mn}(r|r_0)}^2}{k_{y|mn}^2}\right\} \\
      \overline{\abs{R(r)}^2} &= \abs{2\sqrt{\frac{k_{y|mn}}{\DD_{mn}}}\[k_{r|mn}r\]^{-\half}}^2\overline{\cos^2\[\int_{r_{\mathrm{ref}}}^rk_{r|mn}(\rho)\d\rho+\Phi\]} \\
      &= \frac{2k_{y|mn}}{\DD_{mn}k_{r|mn}r} \\
      &= \frac{2\cot(\alpha_{mn})}{\DD_{mn}r} \\
      \mathscr{P}_{\mathrm{TL}} &= S^2\sum_m\abs{\Theta_m(\theta|\theta_0)}^2\left\{\frac{1}{4}\sum_{n=1}^{\inf}\frac{1}{k_{y|mn}^2}\frac{2\cot(\alpha_{0|mn})}{\DD_{mn}r_0}\frac{2\cot(\alpha_{mn})}{\DD_{mn}r}\right\} \\
      &= S^2\sum_m\abs{\Theta_m(\theta|\theta_0)}^2\left\{\frac{1}{r_0r}\sum_{n=1}^{\inf}\frac{\cot(\alpha_{0|mn})\cot(\alpha_{mn})}{k_{y|mn}^2\DD_{mn}^2}\times\Delta n\right\} \\
  \end{split}\end{equation}

  Within the horizontal problem, we have the ray-specific propagation angle, $\alpha$, which is directly tied to the constancy of $k_y$.  This produces a Snell's Law analogue in the horizontal problem, and the derivative of which provides a transform from wavenumber to propagation angle.

  \begin{equation}\begin{split}
      k_{ry}\cos\alpha &= k_{ry}^{\mathrm{min}}\cos\alpha^{\mathrm{min}} = k_{ry}^{\mathrm{src}}\cos\alpha^{\mathrm{src}} \\
      k_y &= k_{ry}\cos\alpha \\
      \dd{k_y}{\alpha} &= -k_{ry}\sin\alpha \\
  \end{split}\end{equation}

  We also need the modal separation, $\dn/\d k_y$, which we get from taking the derivative of the \emph{closed phase integral} \cite{harrison2013ray}.

  \begin{equation}\begin{split}
      n\pi + \phi' + \phi'' &= \int_{r'}^{r''}k_r(\rho)\d\rho \\
      n &= \frac{1}{\pi}\[\int_{r'}^{r''}\sqrt{k_{ry}^2(\rho)-k_y^2}\d\rho - \phi' - \phi''\] \\
      \dd{n}{k_y} &= \frac{1}{\pi}\frac{1}{2}\int_{r'}^{r''}\[k_{ry}^2(\rho)-k_y^2\]^{-\half}\times\(-2k_y\)\d\rho \\
      &= \frac{-k_y}{\pi}\int_{r'}^{r''}\frac{1}{k_r}\d\rho \\
      &= \frac{-k_y}{\pi}\mathfrak{L} \\
      &= \frac{-k_y}{\pi}\frac{\DD}{2k_y} \\
      \dd{n}{k_y} &= \frac{-\DD}{2\pi} \\
  \end{split}\end{equation}

  We now convert the summation over $n$ into an integration over $\dn$ and then map to propagation angle $\alpha$ \cite{harrison2013ray}. We have the choice of whether to define our differential element at the source or receiver.  The modenumber is the same from source to receiver, but the wavenumbers and propagation angles are different when evaluated at the source or receiver location.  By choosing the differential element to be defined at the receiver, future numerical implementation is better suited to defining the angular grid resolution at the receiver locations where the field is being calculated. We will also drop the $mn$ subscripts for clarity since we are no longer concerned with counting the modenumbers.

  \begin{equation}\begin{split}
      \mathscr{P}_{\mathrm{TL}} &= S^2\sum_m\abs{\Theta_m(\theta|\theta_0)}^2\left\{\frac{1}{r_0r}\int_n\frac{\cot(\alpha_0)\cot(\alpha)}{k_y^2\DD^2}\dn\right\} \\
      &= S^2\sum_m\abs{\Theta_m(\theta|\theta_0)}^2\left\{\frac{1}{r_0r}\int_0^{\alpha_{\mathrm{max}}}\frac{\cot(\alpha_0)\cot(\alpha)}{k_y^2\DD^2}\(\dd{n}{k_y}\)\(\dd{k_y}{\alpha}\)\d\alpha\right\} \\
      &= S^2\sum_m\abs{\Theta_m(\theta|\theta_0)}^2\left\{\frac{1}{r_0r}\int_0^{\alpha_{\mathrm{max}}}\frac{\cot(\alpha_0)\cot(\alpha)}{k_y^2\DD^2}\(\frac{-\DD}{2\pi}\)\(-k_{ry}\sin\alpha\)\d\alpha\right\} \\
      &= S^2\sum_m\abs{\Theta_m(\theta|\theta_0)}^2\left\{\frac{1}{2\pi r_0r}\int_0^{\alpha_{\mathrm{max}}}\frac{k_{ry}\cot(\alpha_0)\sin\alpha\cot(\alpha)}{k_y^2\DD}\d\alpha\right\} \\
      &= S^2\sum_m\abs{\Theta_m(\theta|\theta_0)}^2\left\{\frac{1}{2\pi r_0r}\int_0^{\alpha_{\mathrm{max}}}\frac{\cot(\alpha_0)\cos(\alpha)}{k_y\DD\cos(\alpha)}\d\alpha\right\} \\
      &= S^2\sum_m\abs{\Theta_m(\theta|\theta_0)}^2\left\{\frac{1}{2\pi r_0r}\int_0^{\alpha_{\mathrm{max}}}\frac{\cot(\alpha_0)}{k_y\DD}\d\alpha\right\} \\
      &= \sum_m\abs{\Theta_m(\theta|\theta_0)}^2\left\{\frac{8\pi}{r_0r}\int_0^{\alpha_{\mathrm{max}}}\frac{\cot(\alpha_0)}{k_y\DD}\d\alpha\right\} \\
  \end{split}\end{equation}

  This quantity is the incoherent pressure-squared constructed as an incoherent mode sum in the angular coordinate with an incoherent energy flux solution for the horizontally-dependent modal coefficient for each modenumber $m$. The next step is to derive a reflection coefficient for the frontal interface.  We do this by assuming WKB normal modes in the radial direction and satisfying the smoothness boundary conditions at the frontal interface \cite{lin2012analytical}.  Since the front is perpendicular to $r$ and our solution is separable, we only need to work with the terms and factors that depend on the $r$-coordinate.  The appropriate boundary conditions at the frontal interface $r=r_{\mathrm{int}}$ are continuity and smoothness.

  \newcommand{\rint}{r_{\mathrm{int}}}

  \begin{equation}\begin{split}
      p = S\sum_m\Theta_m(\theta_0)\Theta_m(\theta)&\left\{\sum_n\frac{-ie^{ik_{y|mn}\abs{y}}}{2k_{y|mn}}R_{mn}(r_0)R_{mn}(r)\right\} \\
      \evalat{p^<(r)}{r\raro \rint^-} &= \evalat{p^>(r)}{r=\rint^+} \\
      \evalat{\dd{p^<(r)}{r}}{r\raro \rint^-} &= \evalat{\dd{p^>(r)}{r}}{r\raro \rint^-} \\
  \end{split}\end{equation}

  All terms that do not depend on $r$ and operators orthogonal to $r$ pass through these evaluations on both the left and right hand side and thus cancel each other out.  For each $m$ and $n$, we must satisfy the following.

  \begin{equation}\begin{split}
      \evalat{R^<(r)}{r\raro \rint^-} &= \evalat{R^>(r)}{r\raro \rint^+} \\
      \evalat{\d_rR^<(r)}{r\raro \rint^-} &= \evalat{\d_rR^>(r)}{r\raro \rint^+} \\
  \end{split}\end{equation}

  At the present moment, the WKB modefunctions are most useful in their complex exponential form.  We construct the WKB solution as upward and downward propagating plane waves and assign the reflection coefficient to the reflected wave.

  \begin{equation}\begin{split}
      \Let \Phi(r) &= \int_{r_{\mathrm{ref}}}^rk_r(\rho)\d\rho \\
      \dd{\Phi(r)}{r} &= \d_r\int_{r_{\mathrm{ref}}}^rk_r(\rho)\d\rho \\
      \text{Leibniz differentiation under the integral} \\
      \dd{\Phi(r)}{r} &= k_r(r) \\
      R(r) &\sim \CC_1\[k_rr\]^{-\half}e^{+i\Phi(r)} + \CC_2\[k_rr\]^{-\half}e^{-i\Phi(r)} \\
      R^<(r) &= \[k_r^<r\]^{-\half}e^{+i\Phi^<(r)} + \mathfrak{R}\[k_r^<r\]^{-\half}e^{-i\Phi^<(r)} \\
      R^>(r) &= \mathfrak{T}\[k_r^>r\]^{-\half}e^{+i\Phi^>(r)} \\
  \end{split}\end{equation}

  \begin{equation}\begin{split}
      \d_rR^<(r) &= \d_r\(\[k_r^<r\]^{-\half}\)\[e^{+i\Phi^<}+\mathfrak{R}e^{-i\Phi^<}\] + \[k_r^<r\]^{-\half}\d_r\(e^{+i\Phi^<}+\mathfrak{R}e^{-i\Phi^<}\) \\
      &= \frac{-1}{2}\[k_r^<r\]^{-\threehalf}\(k_r^<+r\d_rk_r^<\)\[e^{+i\Phi^<}+\mathfrak{R}e^{-i\Phi^<}\] + \[k_r^<r\]^{-\half}\(ik_r^<e^{+i\Phi^<}-i\mathfrak{R}k_r^<e^{-i\Phi^<}\) \\
      &= \frac{-1}{2}\[k_r^<r\]^{-\threehalf}\(k_r^<+r\d_rk_r^<\)e^{+i\Phi^<} + \frac{-1}{2}\mathfrak{R}\[k_r^<r\]^{-\threehalf}\(k_r^<+r\d_rk_r^<\)e^{-i\Phi^<} \\
      &\qquad\qquad + \[k_r^<r\]^{-\half}(ik_r^<)e^{+i\Phi^<} - \mathfrak{R}(ik_r^<)\[k_r^<r\]^{-\half}e^{-i\Phi^<} \\
      \d_rR^>(r) &= \d_r\(\[k_r^>r\]^{-\half}\)\[\mathfrak{T}e^{+i\Phi^>}\] + \[k_r^>r\]^{-\half}\d_r\(\mathfrak{T}e^{+i\Phi^>}\) \\
      &= \frac{-1}{2}\mathfrak{T}\[k_r^>r\]^{-\threehalf}\(k_r^>+r\d_rk_r^>\)e^{+i\Phi^>} + \mathfrak{T}\[k_r^>r\]^{-\half}(ik_r^>)e^{+i\Phi^>} \\
  \end{split}\end{equation}

  Note that when we evaluate these expressions and take the limit as $r\raro r_{\mathrm{int}}^{\pm}$, $r_{\mathrm{int}}^+=r_{\mathrm{int}}^-$ and the side of $r_{\mathrm{int}}$ to evaluate on is explicit by the definition of $k_r^{</>}$, $R^{</>}$, and $\mathscr{E}^{</>}$.  With the understanding that everything is evaluated at the interface and functions are clearly distinguished as to which side they belong, we will drop the explicit argument of $(r_{\mathrm{int}})$ in the notation. In addition we will use the shorthand, $\mathscr{E}^{</>}_{+\-}$, to replace the complex exponentials.  Immediately we can get rid of some terms by invoking the validity condition for WKB, which states that the change in the wavenumber over a cycle distance is small compared to the magnitude of the wavenumber itself \cite{jensen2011computational, bender2013advanced}.

  \newcommand{\Epl}{\mathscr{E}_+^<}
  \newcommand{\Epr}{\mathscr{E}_+^>}
  \newcommand{\Eml}{\mathscr{E}_-^<}
  \newcommand{\Emr}{\mathscr{E}_-^>}

  \begin{equation}\begin{split}
      \text{WKB condition:}\qquad \frac{\d_rk_r(r)}{k_r(r)} \ll 1 \\
      \(k_r+r\d_rk_r\) &= k_rr\(r^{-1}+\cancel{\frac{\d_rk_r}{k_r}}\) \\
      &= k_r \\
  \end{split}\end{equation}

  \begin{equation}\begin{split}
      \Let \gamma_{</>} = \sqrt{k_r^{</>}r} \\
      \Let \mathscr{E}_{\pm}^{</>} &= e^{\pm i\Phi^{</>}} \\
      \d_rR^<(r) &= \frac{-1}{2}\[k_r^<r\]^{-\threehalf}k_r^<\Epl + \frac{-1}{2}\mathfrak{R}\gamma_<^{-3}k_r^<\Eml + \gamma_<^{-1}(ik_r^<)\Epl - \mathfrak{R}(ik_r^<)\gamma_<^{-1}\Eml \\
      \d_rR^>(r) &= \frac{-1}{2}\mathfrak{T}\gamma_>^{-3}k_r^>\Epr + \mathfrak{T}\gamma_>^{-1}(ik_r^>)\Epr \\
  \end{split}\end{equation}

  \begin{equation}\begin{split}
      \text{Continuity:} \\
      \evalat{r^{-1}R^<(r)}{r\raro \rint^-} &= \evalat{r^{-1}R^>(r)}{r\raro \rint^+} \\
      \bevalat{r^{-1}\gamma_<^{-1}\(\Epl + \mathfrak{R}\; \Eml\)}{r\raro \rint^-} &= \bevalat{r^{-1}\gamma_>^{-1}\(\mathfrak{T}\Epr\)}{r\raro \rint^+} \\
      \gamma_<^{-1}\Epl+\mathfrak{R}\gamma_<^{-1}\Eml &= \mathfrak{T}\gamma_>^{-1}\Epr \\
      \mathfrak{T} &= \frac{\gamma_<^{-1}\Epl+\mathfrak{R}\gamma_<^{-1}\Eml}{\gamma_>^{-1}\Epr}
  \end{split}\end{equation}

  \begin{equation}\begin{split}
      \text{Smoothness:} \\
      \bevalat{\d_rR^<(r) - r^{-1}R^<(r)}{r\raro \rint^-} &= \bevalat{\d_rR^>(r) - r^{-1}R^>(r)}{r\raro \rint^+} \\
  \end{split}\end{equation}

  \begin{multline}
    \frac{-1}{2}\gamma_<^{-3}k_r^<\Epl + \gamma_<^{-1}(ik_r^<)\Epl + \frac{-1}{2}\mathfrak{R}\gamma_<^{-3}k_r^<\Eml - \mathfrak{R}(ik_r^<)\gamma_<^{-1}\Eml - r^{-1}\gamma_<^{-1}\Epl - r^{-1}\mathfrak{R}\gamma_<^{-1}\Eml \\
    = \frac{-1}{2}\mathfrak{T}\gamma_>^{-3}k_r^>\Epr + \mathfrak{T}\gamma_>^{-1}(ik_r^>)\Epr - r^{-1}\mathfrak{T}\gamma_>^{-1}\Epr \\
  \end{multline}

  \begin{multline}
    \frac{-1}{2}\gamma_<^{-3}k_r^<\Epl + \gamma_<^{-1}(ik_r^<)\Epl + \frac{-1}{2}\mathfrak{R}\gamma_<^{-3}k_r^<\Eml - \mathfrak{R}(ik_r^<)\gamma_<^{-1}\Eml - r^{-1}\gamma_<^{-1}\Epl - r^{-1}\mathfrak{R}\gamma_<^{-1}\Eml \\
    = \frac{-1}{2}\frac{\gamma_<^{-1}\Epl+\mathfrak{R}\gamma_<^{-1}\Eml}{\cancel{\gamma_>^{-1}}\cancel{\Epr}}\gamma_>^{-\cancel{3}2}k_r^>\cancel{\Epr} + \frac{\gamma_<^{-1}\Epl+\mathfrak{R}\gamma_<^{-1}\Eml}{\cancel{\gamma_>^{-1}}\cancel{\Epr}}\cancel{\gamma_>^{-1}}(ik_r^>)\cancel{\Epr} - r^{-1}\frac{\gamma_<^{-1}\Epl+\mathfrak{R}\gamma_<^{-1}\Eml}{\cancel{\gamma_>^{-1}}\cancel{\Epr}}\cancel{\gamma_>^{-1}}\cancel{\Epr} \\
  \end{multline}

  \begin{multline}
    \frac{-1}{2}\gamma_<^{-3}k_r^<\Epl + \gamma_<^{-1}(ik_r^<)\Epl + \frac{-1}{2}\mathfrak{R}\gamma_<^{-3}k_r^<\Eml - \mathfrak{R}(ik_r^<)\gamma_<^{-1}\Eml - r^{-1}\gamma_<^{-1}\Epl - r^{-1}\mathfrak{R}\gamma_<^{-1}\Eml \\
    = \frac{-1}{2}\gamma_<^{-1}\gamma_>^{-2}k_r^>\Epl+\frac{-1}{2}\mathfrak{R}\gamma_<^{-1}\gamma_>^{-2}k_r^>\Eml + \gamma_<^{-1}(ik_r^>)\Epl + \mathfrak{R}\gamma_<^{-1}(ik_r^>)\Eml - r^{-1}\gamma_<^{-1}\Epl - r^{-1}\mathfrak{R}\gamma_<^{-1}\Eml \\
  \end{multline}

  \begin{multline}
    \frac{-1}{2}\gamma_<^{-2}k_r^<\Epl + (ik_r^<)\Epl + \frac{-1}{2}\mathfrak{R}\gamma_<^{-2}k_r^<\Eml - \mathfrak{R}(ik_r^<)\Eml - r^{-1}\Epl - r^{-1}\mathfrak{R}\Eml \\
    = \frac{-1}{2}\gamma_>^{-2}k_r^>\Epl+\frac{-1}{2}\mathfrak{R}\gamma_>^{-2}k_r^>\Eml + (ik_r^>)\Epl + \mathfrak{R}(ik_r^>)\Eml - r^{-1}\Epl - r^{-1}\mathfrak{R}\Eml \\
  \end{multline}

  \begin{multline}
    \mathfrak{R}\[\frac{-1}{2}\gamma_<^{-2}k_r^< - (ik_r^<) - \cancel{r^{-1}} + \frac{+1}{2}\gamma_>^{-2}k_r^> - (ik_r^>) + \cancel{r^{-1}}\]\Eml \\
    = \[\frac{-1}{2}\gamma_>^{-2}k_r^> + (ik_r^>) - \cancel{r^{-1}} - \frac{-1}{2}\gamma_<^{-2}k_r^< - (ik_r^<) + \cancel{r^{-1}}\]\Epl \\
  \end{multline}

  \begin{equation}\begin{split}
      \mathfrak{R} &= \frac{\frac{-1}{2}\gamma_>^{-2}k_r^> + (ik_r^>) - \frac{-1}{2}\gamma_<^{-2}k_r^< - (ik_r^<)}{\frac{-1}{2}\gamma_<^{-2}k_r^< - (ik_r^<) + \frac{+1}{2}\gamma_>^{-2}k_r^> - (ik_r^>)}\mathscr{E}_{+2}^< \\
      &= \frac{\frac{-1}{2}\[k_r^>r\]^{-1}k_r^> + (ik_r^>) - \frac{-1}{2}\[k_r^<r\]^{-1}k_r^< - (ik_r^<)}{\frac{-1}{2}\[k_r^<r\]^{-1}k_r^< - (ik_r^<) + \frac{+1}{2}\[k_r^>r\]^{-1}k_r^> - (ik_r^>)}\mathscr{E}_{+2}^< \\
      &= \frac{\cancel{\frac{-1}{2}r^{-1}} + (ik_r^>) + \cancel{\frac{1}{2}r^{-1}} - (ik_r^<)}{\cancel{\frac{-1}{2}r^{-1}} - (ik_r^<) + \cancel{\frac{1}{2}r^{-1}} - (ik_r^>)}\mathscr{E}_{+2}^< \\
      \mathfrak{R} &= \frac{k_r^<-k_r^>}{k_r^<+k_r^>}\times\mathscr{E}_{+2}^< \\
      \left\{\mathcal{RL}\right\} &= \abs{\mathfrak{R}}^{2\abs{y}/\DD} \\
  \end{split}\end{equation}

  This looks like the standard form for a Rayleigh reflection coefficient, but we are using $k_r$ instead of the characteristic impedance.  For our model, we neglect the phase shift in the reflection coefficient, and model the reflection loss as a gradual loss over $|y|/\DD$.

  \begin{equation}\begin{split}
      \mathscr{P}_{\mathrm{TL}} &= \sum_m\abs{\Theta_m(\theta|\theta_0)}^2\left\{\frac{8\pi}{r_0r}\int_0^{\alpha_{\mathrm{max}}}\abs{\frac{k_r^<-k_r^>}{k_r^<+k_r^>}}^{2\abs{y}/\DD}\frac{\cot(\alpha_0)}{k_y\DD}\d\alpha\right\} \\
  \end{split}\end{equation}

  Now to derive the \emph{convergence factor} we return to the coherent mode sum for the horizontal problem and retain the cross products \cite{harrison2013ray}.

  \begin{equation}\begin{split}
      \mathscr{P}_{\mathrm{TL}} &= |p|^2 \\
      &= S^2\sum_m\abs{\Theta_m(\theta|\theta_0)}^2\abs{B_m}^2 \\
      \abs{B_m}^2 &= \frac{1}{4}\(\sum_{n=1}^\inf\frac{\abs{R_{mn}(r|r_0)}^2}{k_{y|mn}^2} + \sum_{n=2}^\inf\sum_{q=1}^{n-1}\frac{R_{mn}(r|r_0)R_{mq}^*(r|r_0)}{k_{y|mn}k_{y|mq}} 2\cos\[(k_{y|mn}-k_{y|mq})\abs{y}\]\) \\
      &= \abs{B_m}^2_I + \abs{B_m}^2_C \\
  \end{split}\end{equation}

  Considering only the coherent term, we insert WKB modefunctions, now expressing the oscillation in terms of a sine function with some other phase offset $\vphi$ \cite{harrison2013ray}.

  \begin{equation}\begin{split}
      \Let \Phi_{mn}(r) &= \int_{r_{\mathrm{ref}}}^rk_{r|mn}(\rho)\d\rho + \vphi \\
      \abs{B_m}^2_C &= \frac{1}{4}\sum_{n=2}^\inf\sum_{q=1}^{n-1}\frac{2\cos\[(k_{y|mn}-k_{y|mq})\abs{y}\]}{k_{y|mn}k_{y|mq}}R_{mn}(r|r_0)R_{mq}^*(r|r_0) \\
      R_{mn}(r) &= 2\sqrt{\frac{k_{y|mn}}{\DD_{mn}}}\[k_{r|mn}r\]^{-\half}\sin\[\Phi_{mn}(r)\] \\
      R_{mn}(r|r_0) &= 4\(\frac{k_{y|mn}}{\DD_{mn}}\)\[k_{r_0|mn}r_0k_{r|mn}r\]^{-\half}\sin\[\Phi_{mn}(r)\]\sin\[\Phi_{mn}(r_0)\] \\
      \abs{B_m}^2_C &= \frac{1}{4}\sum_{n=2}^\inf\sum_{q=1}^{n-1}\frac{2\cos\[(k_{y|mn}-k_{y|mq})\abs{y}\]}{k_{y|mn}k_{y|mq}} \\
      &\qquad\qquad \times 4\(\frac{k_{y|mn}}{\DD_{mn}}\)\[k_{r_0|mn}r_0k_{r|mn}r\]^{-\half}\sin\[\Phi_{mn}(r)\]\sin\[\Phi_{mn}(r_0)\] \\
      &\qquad\qquad \times 4\(\frac{k_{y|mq}}{\DD_{mq}}\)\[k_{r_0|mq}r_0k_{r|mq}r\]^{-\half}\sin\[\Phi_{mq}(r)\]\sin\[\Phi_{mq}(r_0)\] \\
  \end{split}\end{equation}

  \begin{equation}\begin{split}
      \abs{B_m}^2_C &= \frac{1}{4}\frac{32}{r_0r}\sum_{n=2}^\inf\sum_{q=1}^{n-1}\frac{\cos\[(k_{y|mn}-k_{y|mq})\abs{y}\]}{k_{y|mn}k_{y|mq}} \\
      &\qquad\qquad \times \(\frac{k_{y|mn}}{\DD_{mn}}\)\[k_{r_0|mn}k_{r|mn}\]^{-\half}\sin\[\Phi_{mn}(r)\]\sin\[\Phi_{mn}(r_0)\] \\
      &\qquad\qquad \times \(\frac{k_{y|mq}}{\DD_{mq}}\)\[k_{r_0|mq}k_{r|mq}\]^{-\half}\sin\[\Phi_{mq}(r)\]\sin\[\Phi_{mq}(r_0)\] \\
  \end{split}\end{equation}

  \begin{equation}\begin{split}
      2\sin\[\Phi_n\]\sin\[\Phi_q\] &= \cos\[\Phi_n-\Phi_q\]-\cancel{\cos\[\Phi_n+\Phi_q\]} \\
      \Raro\qquad \abs{B_m}^2_C &= \frac{1}{4}\frac{32}{r_0r}\sum_{n=2}^\inf\sum_{q=1}^{n-1}\frac{\(\frac{k_{y|mn}}{\DD_{mn}}\)\(\frac{k_{y|mq}}{\DD_{mq}}\)}{k_{y|mn}k_{y|mq}}\[k_{r_0|mn}k_{r|mn}k_{r_0|mq}k_{r|mq}\]^{-\half} \\
      &\qquad \times \frac{1}{4}\cos\[(k_{y|mn}-k_{y|mq})\abs{y}\]\cos\[\Phi_{mn}(r)-\Phi_{mq}(r)\]\cos\[\Phi_{mn}(r_0)-\Phi_{mq}(r_0)\] \\
      &=  \frac{1}{4}\frac{32}{r_0r}\sum_{n=2}^\inf\sum_{q=1}^{n-1}\frac{\[k_{r_0|mn}k_{r|mn}k_{r_0|mq}k_{r|mq}\]^{-\half}}{\DD_{mn}\DD_{mq}} \\
      &\qquad \times \frac{1}{4}\cos\[(k_{y|mn}-k_{y|mq})\abs{y}\]\cos\[\Phi_{mn}(r)-\Phi_{mq}(r)\]\cos\[\Phi_{mn}(r_0)-\Phi_{mq}(r_0)\] \\
  \end{split}\end{equation}

  \begin{equation}\begin{split}
      \cos(a)\times\cos(b)\times\cos(c) &= \frac{1}{4}\sum_4\cos\[a \pm b \pm c\] \\
      \Raro\qquad \abs{B_m}^2_C &=  \frac{1}{4}\frac{32}{r_0r}\sum_{n=2}^\inf\sum_{q=1}^{n-1}\frac{\[k_{r_0|mn}k_{r|mn}k_{r_0|mq}k_{r|mq}\]^{-\half}}{\DD_{mn}\DD_{mq}} \\
      &\qquad \times \frac{1}{16}\sum_4\cos\[(k_{y|mn}-k_{y|mq})\abs{y} \pm \(\Phi_{mn}(r)-\Phi_{mq}(r)\) \pm \(\Phi_{mn}(r_0)-\Phi_{mq}(r_0)\)\] \\
  \end{split}\end{equation}

  Since we are primarily concerned with the interference from neighboring modes, we expand these differences in the cosine arguments as Taylor series in $m-n$ centered about $m=n$ \cite{harrison2013ray}.

  \begin{equation}\begin{split}
      \Let j &= n-q \tand \text{Taylor expand  } k_{y|mn} \text{  about  } k_{y|mq} \\
      k_{y|mn} &\approx k_{y|mq} + \dd{k_{y|mn}}{n}j + \mathscr{O}(j^2) \\
      k_{y|mn} - k_{y|mq} &\approx \dd{k_{y|mn}}{n}j \\
      \text{Similarly,}& \\
      \Phi_{mn}(r|r_0) - \Phi_{mq}(r|r_0) &\approx \dd{\Phi_{mn}(r|r_0)}{n}j \\
  \end{split}\end{equation}

  The derivatives in these approximations are simply the modal separation obtained from taking the derivative of the \emph{closed phase integral} and the derivative of the \emph{open phase integral} with respect to $k_y$ \cite{harrison2013ray}.

  \begin{equation}\begin{split}
      n\pi + \phi' + \phi'' &= \int_{r'}^{r''}k_r(\rho)\d\rho \\
      \dd{n}{k_y} &= \frac{-\DD}{2\pi} \\
      \dd{\Phi(r_0)}{k_y} &= \dd{}{k_y} \int_{r_{\mathrm{ref}}}^{r_0}k_r(\rho)\d\rho \\
      &= \dd{}{k_y} \int_{r_{\mathrm{ref}}}^{r_0}\sqrt{k_{ry}^2(\rho)-k_y^2}\d\rho \\
      &= \int_{r_{\mathrm{ref}}}^{r_0}\frac{-k_y}{\sqrt{k_{ry}^2(\rho)-k_y^2}}\d\rho \\
      &= \int_{r_{\mathrm{ref}}}^{r_0}\frac{-k_y}{k_r}\d\rho \\
      &= -\int_{r_{\mathrm{ref}}}^{r_0}\cot(\alpha)\d\rho \\
      &= -\DD_{\mathrm{src}} \\
  \end{split}\end{equation}

  \begin{equation}\begin{split}
      k_{y|mn}-k_{y|mq} &\approx \frac{-2\pi}{\DD_{mn}}j \\
      \Phi_{mn}(r)-\Phi_{mq}(r) &\approx \dd{\Phi_{mn}(r)}{k_{y|mn}}\dd{k_{y|mn}}{n}j \\
      &\approx \DD_{mn}^{\mathrm{rcv}}\cdot\frac{2\pi}{\DD_{mn}}j \\
      \Phi_{mn}(r_0)-\Phi_{mq}(r_0) &\approx \DD_{mn}^{\mathrm{src}}\cdot\frac{2\pi}{\DD_{mn}}j \\
  \end{split}\end{equation}

  \begin{equation}\begin{split}
      \abs{B_m}^2_C &=  \frac{1}{16}\frac{32}{r_0r}\sum_{n=2}^\inf\sum_{q=1}^{n-1}\frac{\[k_{r_0|mn}k_{r|mn}k_{r_0|mq}k_{r|mq}\]^{-\half}}{\DD_{mn}\DD_{mq}} \times \frac{1}{4}\sum_4\cos\[2\pi\(\frac{\abs{y}}{\DD_{mn}} \pm \frac{\DD_{mn}^{\mathrm{rcv}}}{\DD_{mn}} \pm \frac{\DD_{mn}^{\mathrm{src}}}{\DD_{mn}}\) j\] \\
  \end{split}\end{equation}

  Then we rearrange our summation to be performed over $j=n-q$, and assume that coefficients in $n$ and $q$ are roughly equivalent, i.e. a modal continuum of densely packed modes \cite{harrison2013ray}. We convert the summation over $n$ to an integration over $\dn$ and then transform the differential element to propagation angle at the receiver, $\d\alpha$. The transforms make use of the same derivatives of the \emph{closed phase integral} and Snell's Law analog that were used in the incoherent derivation. When incorporating range-dependence in the $y$-direction via the adiabatic approximation (aka Born-Oppenheimer approximation), it's important to remember that the modenumber $n$ is defined independently of source or receiver position but the differential transform factors ($\dn/\d k_y$ and $\d k_y/\d\alpha$) are derived from a specified location, typically at the source or receiver position. We will expand our differential element at the receiver position since defining the angular grids at each receiver works better for guaranteeing adequate angular resolution in a TL computation.

  \newcommand{\pmDD}{\frac{\abs{y}}{\DD_m} \pm \frac{\DD_m^{\mathrm{rcv}}}{\DD_m} \pm \frac{\DD_m^{\mathrm{src}}}{\DD_m}}

  \begin{equation}\begin{split}
      \abs{B_m}^2_C &=  \frac{2}{r_0r}\sum_n\frac{1}{\DD_{mn}^2k_{r_0|mn}k_{r|mn}} \times \frac{1}{4}\sum_4\sum_j\cos\[2\pi\(\pmDD\) j\] \\
      &=  \frac{2}{r_0r}\sum_n\frac{\cot(\alpha_0)\cot(\alpha)}{k_{y|mn}^2\DD_{mn}^2} \times \frac{1}{4}\sum_4\sum_j\cos\[2\pi\(\pmDD\) j\] \\
      &= \frac{2}{r_0r}\int_n\frac{\cot(\alpha_0)\cot(\alpha)}{k_{y|m}^2\DD_{m}^2} \times \frac{1}{4}\sum_4\sum_j\cos\[2\pi\(\pmDD\) j\] \dn \\
      &= \frac{2}{r_0r}\int_0^{\alpha_{\mathrm{max}}}\frac{\cot(\alpha_0)\cot(\alpha)}{k_{y|m}^2\DD_{m}^2} \times \frac{1}{4}\sum_4\sum_j\cos\[2\pi\(\pmDD\) j\]\dd{n}{k_y}\dd{k_y}{\alpha} \d\alpha \\
      &= \frac{2}{r_0r}\int_0^{\alpha_{\mathrm{max}}}\frac{\cot(\alpha_0)\cot(\alpha)}{k_{y|m}^2\DD_{m}^2} \times \frac{1}{4}\sum_4\sum_j\cos\[2\pi\(\pmDD\) j\]\(\frac{-\DD_m}{2\pi}\)\(-k_{ry|m}\sin\alpha\)\d\alpha \\
      &= \frac{1}{\pi r_0r}\int_0^{\alpha_{\mathrm{max}}}\frac{\cot(\alpha_0)\cot(\alpha)k_{ry|m}\sin(\alpha)}{k_{y|m}^2\DD_{m}} \times \frac{1}{4}\sum_4\sum_j\cos\[2\pi\(\pmDD\) j\]\d\alpha \\
      &= \frac{1}{\pi r_0r}\int_0^{\alpha_{\mathrm{max}}}\frac{\cot(\alpha_0)}{k_{y|m}\DD_{m}} \times \frac{1}{4}\sum_4\sum_j\cos\[2\pi\(\pmDD\) j\]\d\alpha \\
  \end{split}\end{equation}

  Now we add back the incoherent solution for a total expression of $|B_m|^2$.

  \begin{equation}\begin{split}
      \abs{B_m}^2_I &= \frac{1}{2\pi r_0r}\int_0^{\alpha_{\mathrm{max}}}\frac{\cot(\alpha_0)}{k_y\DD}\d\alpha \\
      \abs{B_m}^2 &= \abs{B_m}^2_I + \abs{B_m}^2_C \\
      &= \frac{1}{2\pi r_0r}\int_0^{\alpha_{\mathrm{max}}}\frac{\cot(\alpha_0)}{k_y\DD}\d\alpha \\
      &\qquad + \frac{1}{\pi r_0r}\int_0^{\alpha_{\mathrm{max}}}\frac{\cot(\alpha_0)}{k_{y|m}\DD_{m}} \times \frac{1}{4}\sum_4\sum_j\cos\[2\pi\(\pmDD\) j\]\d\alpha \\
      &= \frac{1}{2\pi r_0r}\int_0^{\alpha_{\mathrm{max}}}\frac{\cot(\alpha_0)}{k_{y|m}\DD_m}\times\frac{1}{4}\sum_4\left\{1+2\sum_{j=1}^N\cos\[2\pi\(\pmDD\)j\]\right\}\d\alpha \\
  \end{split}\end{equation}

  In Harrison's 2013 paper, a smoothed approximation of this cosine series is given \cite{harrison2013ray}.

  \begin{equation}\begin{split}
      1+2\sum_{j=1}^N\cos(Xj) &\approx (1+2N)\times\exp\[\frac{-(2N+1)^2}{\pi}\sin^2(X/2)\] \\
      1+2\sum_{j=1}^N\cos\[2\pi\(\pmDD\)j\] & \\
      \approx (1+2N)\times&\exp\[\frac{-(2N+1)^2}{\pi}\sin^2\(\pi\(\pmDD\)\)\] \\
  \end{split}\end{equation}

  Thus we have derived both a frontal reflection loss and the convergence factor for this hybrid normal mode and energy flux model.

  \begin{equation}\begin{split}
      \Theta_m(\theta) &= \sqrt{\frac{2}{\theta_B}}\sin(\vkap_m\theta) \twhere \vkap_m = (n-\half)\pi/\theta_B \\
      \left\{\mathcal{RL}\right\} &= \abs{\frac{k_r^<(\rint)-k_r^>(\rint)}{k_r^<(\rint)+k_r^>(\rint)}}^{2\abs{y}/\DD_m} \\
      \left\{\mathcal{CF}\right\} &= \frac{1}{4}\sum_4\left\{(1+2N)\exp\[\frac{-(1+2N)^2}{\pi}\sin^2\(\pi\[\pmDD\]\)\]\right\} \\
      \abs{B_m}^2 &= \frac{1}{2\pi r_0r}\int_0^{\alpha_{\mathrm{max}}}\left\{\mathcal{RL}\right\}\times\frac{\cot(\alpha_0)}{k_{y|m}\DD_m}\times\left\{\mathcal{CF}\right\}\d\alpha \\
      \mathscr{P}_{\mathrm{TL}} &= S^2\sum_m\abs{\Theta_m(\theta)}^2\abs{\Theta_m(\theta_0)}^2\abs{B_m}^2 \\
      &= \sum_m\abs{\Theta_m(\theta)}^2\abs{\Theta_m(\theta_0)}^2 \left\{\frac{8\pi}{r_0r}\int_0^{\alpha_{\mathrm{max}}}\left\{\mathcal{RL}\right\}\times\frac{\cot(\alpha_0)}{k_{y|m}\DD_m}\times\left\{\mathcal{CF}\right\}\d\alpha\right\} \\
      \mathrm{TL} &= -10\log_{10}\[\frac{\rho_0c_0}{\rho c}\mathscr{P}_{\mathrm{TL}}\] \\
  \end{split}\end{equation}

  This model could be further modified to account for adiabatic range-dependence in the $y$-direction \cite{harrison2015efficient}. If there were adiabatic axial range dependence, then the propagation angles are mapped between source and receiver axial ranges by use of the \emph{ray invariant}, which is invariant under the WKB and adiabatic approximations. The adiabatic modes approximation neglects the coupling matrices that come from the horizontal derivative of the vertical modes \cite{jensen2011computational,pierce1965extension,weinberg1974horizontal}.

  \begin{equation}\begin{split}
      \lap p + k^2 p &= 0 \\
      \Let \lap &= r^{-1}\p_r(r\p_r) + r^{-2}\p_\theta^2 + \p_y^2 \\
      \Ansatz p &= \sum_mB_m(r,y)\Theta_m(\theta|y) \\
  \end{split}\end{equation}

  \begin{equation}\begin{split}
      r^{-1}\p_r\(r\p_r\sum_mB_m\Theta_m\) + r^{-2}\p_\theta^2\(\sum_mB_m\Theta_m\) + \p_y^2\(\sum_mB_m\Theta_m\) + k^2\(\sum_mB_m\Theta_m\) &= 0 \\
      \sum_m\[\Theta_m\(r^{-1}\p_rB_m + \p_r^2B_m\) + r^{-2}B_m\p_\theta^2\(\Theta_m\)\right. \qquad\qquad\qquad\qquad\qquad\qquad & \\
      \left. + \(\Theta_m\p_y^2B_m+2\p_y\Theta_m\p_yB_m+B_m\p_y^2\Theta_m\) + k^2 B_m\Theta_m\] &= 0 \\
  \end{split}\end{equation}

  We multiply this equation by $\Theta_n$ and integrate over the domain of $\theta$. Note that $\vkap_m(y)$ is dependent on axial range.

  \begin{equation}\begin{split}
      \int_0^{\theta_B}\sum_m\left\{\begin{matrix}\\\end{matrix}\Theta_m\(r^{-1}\p_rB_m + \p_r^2B_m\) - \frac{\vkap^2}{r^2}B_m\Theta_m \right. \qquad\qquad\qquad\qquad\qquad\qquad & \\
      \left. + \(\Theta_m\p_y^2B_m+2\p_y\Theta_m\p_yB_m+B_m\p_y^2\Theta_m\) + \begin{matrix}\\\end{matrix}k^2 B_m\Theta_m\right\}\Theta_n\d\theta &= 0 \\
      \sum_m\left\{\begin{matrix}\\\end{matrix}\(r^{-1}\p_rB_m + \p_r^2B_m\)\int_0^{\theta_B}\Theta_m\Theta_n\d\theta +\(k^2 -  \frac{\vkap^2}{r^2}\)B_m\int_0^{\theta_B}\Theta_m\Theta_n\d\theta \right. \qquad\qquad\qquad\qquad & \\
      \left. + \(\p_y^2B_m\int_0^{\theta_B}\Theta_m\Theta_n\d\theta+2\p_yB_m\int_0^{\theta_B}(\p_y\Theta_m)\Theta_n\d\theta+B_m\int_0^{\theta_B}(\p_y^2\Theta_m)\Theta_n\d\theta\) \begin{matrix}\\\end{matrix}\right\} &= 0 \\
      \sum_m\left\{\begin{matrix}\\\end{matrix}\(r^{-1}\p_rB_m + \p_r^2B_m + \p_y^2B_m + \(k^2 -  \frac{\vkap^2}{r^2}\)B_m\)\delta_{mn} \right. \qquad\qquad\qquad\qquad\qquad\qquad & \\
      \left. + \(2\p_yB_m\int_0^{\theta_B}(\p_y\Theta_m)\Theta_n\d\theta+B_m\int_0^{\theta_B}(\p_y^2\Theta_m)\Theta_n\d\theta\) \begin{matrix}\\\end{matrix}\right\} &= 0 \\
      \begin{matrix}\\\end{matrix}\(r^{-1}\p_rB_n + \p_r^2B_n + \p_y^2B_n + \(k_n^2 -  \frac{\vkap_n^2}{r^2}\)B_n\) \qquad\qquad\qquad\qquad\qquad\qquad & \\
      = -\sum_m\(2\p_yB_m\int_0^{\theta_B}(\p_y\Theta_m)\Theta_n\d\theta+B_m\int_0^{\theta_B}(\p_y^2\Theta_m)\Theta_n\d\theta\) & \\
  \end{split}\end{equation}

  We have the adiabatic differential equation defining $B_n$ if we assume that all terms on the right hand side involving horizontal derivatives of $\Theta$ are zero \cite{jensen2011computational}.  One way of accomplishing this is by assuming that $\p_y\Theta\ll 1$. This implies that the solution to the horizontal coefficient is independently calculated at each horizontal position and does not depend on the solution at any other horizontal position. Each mode retains a unique identity corresponding to its modenumber and to its ray invariant.

  The ray invariant is a cycle calculation related to the closed phase integral and the cycle distance, that can be interpreted as the time taken for the vertical phase to traverse a cycle in the watercolumn. However for this horizontal problem, it would be the time taken for the radial phase to traverse a cycle in the horizontal problem \cite{weston1959guided,weston1980acoustic1,harrison2015efficient}.

  \begin{equation}\begin{split}
      \TT &= \int_{r'}^{r''}\frac{\sin\alpha}{c}\dr \\
  \end{split}\end{equation}

  With range dependence in the axial direction, we have to distinguish the evaluation position of certain terms in the derivation since they are no longer the same at source and receiver axial ranges.  The wavenumbers $k_{ry}$, $k_y$, and $k_r$, and the cycle distance $\DD$ can all be calculated at the source or receiver position.  These quantities can all be mapped directly to on-axis propagation angles, which are then mapped to another axial range by use of the ray invariant. Alternatively, you can map all propagation angles to the ray invariant and use the ray invariant to interpolate wavenumbers and cycle distances in the adiabatic approximation. Thus we return to the modesum to track the source/receiver position distinction when substituting in the WKB modefunctions \cite{harrison2015efficient}.

  \begin{equation}\begin{split}
      \NB \DD_{mn} &= 2\int_{r'}^{r''}\frac{k_{y|mn}}{k_{r|mn}}\dr \\
      \DD_{mn0} &= 2\int_{r'}^{r''}\frac{k_{y_0|mn}}{k_{r_0|mn}}\dr \\
      \DD_{mn}^{\mathrm{rcv}}(r) &= \int_{r_{\mathrm{ref}}}^r\frac{k_{y|mn}}{k_{r|mn}}\dr \\
      \DD_{mn}^{\mathrm{src}}(r) &= \int_{r_{\mathrm{ref}}}^{r_0}\frac{k_{y_0|mn}}{k_{r_0|mn}}\dr \\
  \end{split}\end{equation}

  \begin{equation}\begin{split}
      \abs{B_m}^2_I + \abs{B_m}^2_C &= \frac{1}{4}\(\sum_{n=1}^\inf\frac{\abs{R_{mn}(r)}^2\abs{R_{mn}(r_0)}^2}{k_{y|mn}^2} \right. \\
      & \left. + \sum_{n=2}^\inf\sum_{q=1}^{n-1}\frac{R_{mn}(r)R_{mn}(r_0)R_{mq}^*(r)R_{mq}^*(r_0)}{k_{y|mn}k_{y|mq}} 2\cos\[\int_0^{\abs{y}}k_{y|mn}(\gamma)\d\gamma-\int_0^{\abs{y}}k_{y|mq}(\gamma)\d\gamma\]\) \\
      R_{mn}(r) &\sim \frac{2}{\sqrt{r}}\sqrt{\frac{k_{y|mn}}{\DD_{mn}k_{r|mn}}}\cos\[\int_{r_{\mathrm{ref}}}^rk_{r|mn}(\rho)\d\rho+\Phi\] \\
      R_{mn}(r_0) &\sim \frac{2}{\sqrt{r_0}}\sqrt{\frac{k_{y_0|mn}}{\DD_{mn0}k_{r_0|mn}}}\cos\[\int_{r_{\mathrm{ref}}}^rk_{r_0|mn}(\rho)\d\rho+\Phi\] \\
  \end{split}\end{equation}

  \begin{equation}\begin{split}
      \abs{B_m}^2_I &= \frac{4}{r_0r}\sum_{n=1}^\inf\(\frac{k_{y|mn}}{\DD_{mn}k_{r|mn}}\)\(\frac{k_{y0|mn}}{\DD_{mn0}k_{r0|mn}}\)\(\frac{1}{k_{y|mn}^2}\)\cos^2\[...\]\cos^2\[...\]\Delta n \\
      &\sim \frac{1}{r_0r}\sum_{n=1}^\inf\(\frac{k_{y|mn}}{\DD_{mn}k_{r|mn}}\)\(\frac{k_{y0|mn}}{\DD_{mn0}k_{r0|mn}}\)\(\frac{1}{k_{y|mn}^2}\)\Delta n \\
      &\sim \frac{1}{r_0r}\int_0^{\alpha_{\mathrm{max}}}\(\frac{1}{\DD_{m}k_{r|m}}\)\(\frac{k_{y0|m}}{\DD_{m0}k_{r0|m}}\)\(\frac{1}{k_{y|m}}\)\dd{n}{k_{y|m}}\dd{k_{y|m}}{\alpha}\d\alpha \\
      &\sim \frac{1}{r_0r}\int_0^{\alpha_{\mathrm{max}}}\(\frac{1}{\DD_{m}k_{r|m}}\)\(\frac{k_{y0|m}}{\DD_{m0}k_{r0|m}}\)\(\frac{1}{k_{y|m}}\)\(\frac{\DD_{m}}{2\pi}\)\(k_{ry|m}\sin(\alpha)\)\d\alpha \\
      &\sim \frac{1}{2\pi r_0r}\int_0^{\alpha_{\mathrm{max}}}\(\frac{1}{\DD_{m0}}\)\(\frac{k_{y0|m}\cancel{k_{ry|m}}\cancel{k_{r|m}}}{\cancel{k_{r|m}}k_{r0|m}k_{y|m}\cancel{k_{ry|m}}}\)\d\alpha \\
      &\sim \frac{1}{2\pi r_0r}\int_0^{\alpha_{\mathrm{max}}}\frac{\cot(\alpha_0)}{k_{y|m}\DD_{m0}}\d\alpha \\
  \end{split}\end{equation}

  \begin{equation}\begin{split}
      R_{mn}(r) &= \frac{2}{\sqrt{r}}\sqrt{\frac{k_{y|mn}}{\DD_{mn}k_{r|mn}}}\sin\[\Phi_{mn}(r)\] \\
      \abs{B_m}^2_C &= \frac{1}{4}\sum_{n=2}^\inf\sum_{q=1}^{n-1}\frac{R_{mn}(r)R_{mn}(r_0)R_{mq}^*(r)R_{mq}^*(r_0)}{k_{y|mn}k_{y|mq}} 2\cos\[\int_0^{\abs{y}}k_{y|mn}(\gamma)\d\gamma-\int_0^{\abs{y}}k_{y|mq}(\gamma)\d\gamma\] \\
  \end{split}\end{equation}

  \begin{equation}\begin{split}
      R_{mn}(r)R_{mn}(r_0)R_{mq}^*(r)R_{mq}^*(r_0) &= \frac{16}{r_0r}\sqrt{\frac{k_{y|mn}}{\DD_{mn}k_{r|mn}}}\sqrt{\frac{k_{y_0|mn}}{\DD_{mn0}k_{r_0|mn}}}\sqrt{\frac{k_{y|mq}}{\DD_{mq}k_{r|mq}}}\sqrt{\frac{k_{y_0|mq}}{\DD_{mq0}k_{r_0|mq}}} \\
      &\qquad\qquad \times \sin\[\Phi_{mn}(r)\] \sin\[\Phi_{mn}(r_0)\] \sin\[\Phi_{mq}(r)\] \sin\[\Phi_{mq}(r_0)\] \\
      &\approx \frac{16}{r_0r}\sqrt{\frac{k_{y|mn}}{\DD_{mn}k_{r|mn}}\frac{k_{y_0|mn}}{\DD_{mn0}k_{r_0|mn}}\frac{k_{y|mq}}{\DD_{mq}k_{r|mq}}\frac{k_{y_0|mq}}{\DD_{mq0}k_{r_0|mq}}} \\
      &\qquad\qquad \times \frac{1}{4}\cos\[\Phi_{mn}(r)-\Phi_{mq}(r)\]\cos\[\Phi_{mn}(r_0)-\Phi_{mq}(r_0)\] \\
  \end{split}\end{equation}

  \begin{equation}\begin{split}
      \abs{B_m}^2_C &= \frac{1}{2}\sum_{n=2}^\inf\sum_{q=1}^{n-1}\frac{4}{r_0r}\sqrt{\frac{k_{y|mn}}{\DD_{mn}k_{r|mn}}\frac{k_{y_0|mn}}{\DD_{mn0}k_{r_0|mn}}\frac{k_{y|mq}}{\DD_{mq}k_{r|mq}}\frac{k_{y_0|mq}}{\DD_{mq0}k_{r_0|mq}}}\frac{1}{k_{y|mn}k_{y|mq}} \\
      & \qquad \times \cos\[\Phi_{mn}(r)-\Phi_{mq}(r)\]\cos\[\Phi_{mn}(r_0)-\Phi_{mq}(r_0)\]\cos\[\int_0^{\abs{y}}k_{y|mn}(\gamma)\d\gamma-\int_0^{\abs{y}}k_{y|mq}(\gamma)\d\gamma\] \\
      &= \frac{1}{2}\sum_{n=2}^\inf\sum_{q=1}^{n-1}\frac{4}{r_0r}\sqrt{\frac{k_{y|mn}}{\DD_{mn}k_{r|mn}}\frac{k_{y_0|mn}}{\DD_{mn0}k_{r_0|mn}}\frac{k_{y|mq}}{\DD_{mq}k_{r|mq}}\frac{k_{y_0|mq}}{\DD_{mq0}k_{r_0|mq}}}\frac{1}{k_{y|mn}k_{y|mq}} \\
      & \times \frac{1}{4}\sum_4\cos\(\[\Phi_{mn}(r)-\Phi_{mq}(r)\]\pm\[\Phi_{mn}(r_0)-\Phi_{mq}(r_0)\]\pm\[\int_0^{\abs{y}}k_{y|mn}(\gamma)\d\gamma-\int_0^{\abs{y}}k_{y|mq}(\gamma)\d\gamma\]\) \\
  \end{split}\end{equation}

  Approximate phase differences using 1st order Taylor series expansions.

  \begin{equation}\begin{split}
      \Let j &= n - q \\
      \Phi_{mn}(r)-\Phi_{mq}(r) &\approx \dd{\Phi_{mn}(r)}{n} \times j \\
      &= \dd{\Phi_{mn}(r)}{k_{y|mn}}\dd{k_{y|mn}}{n} \times j \\
      &= -2\pi \times \frac{\DD_{mn}^{\mathrm{rcv}}}{\DD_{mn}} \times j \\
      \Phi_{mn}(r_0)-\Phi_{mq}(r_0) &\approx \dd{\Phi_{mn}(r_0)}{n}j \\
      &\approx \dd{\Phi_{mn}(r_0)}{k_{y|mn}}\dd{k_{y|mn}}{n} \times j \\
      &= -2\pi \times \frac{\DD_{mn}^{\mathrm{src}}}{\DD_{mn0}} \times j \\
      \int_0^{\abs{y}}k_{y|mn}(\gamma)\d\gamma-\int_0^{\abs{y}}k_{y|mq}(\gamma)\d\gamma &\approx \int_0^{\abs{y}}\dd{k_{y|mn}(\gamma)}{n}\d\gamma \times j \\
      &= -2\pi \times \int_0^{\abs{y}}\frac{1}{\DD_{mn}(\gamma)}\d\gamma \times j \\
  \end{split}\end{equation}

  \begin{equation}\begin{split}
      \frac{1}{4}\sum_4\[...\] &\approx \frac{1}{4}\sum_4\cos\(-2\pi\[\int_0^{\abs{y}}\frac{\d\gamma}{\DD_{mn}(\gamma)} \pm \frac{\DD_{mn}^{\mathrm{rcv}}}{\DD_{mn}} \pm \frac{\DD_{mn0}^{\mathrm{src}}}{\DD_{mn0}}\]j\) \\
  \end{split}\end{equation}

  With this first-order Taylor series approximation, we treat the coefficient wavenumbers and cycle distances under the radical as if $n=q$ combining these terms.  At the same time we rearrange to a summation over $j$.

  \begin{equation}\begin{split}
      \{\mathcal{CF}_0\} &= \frac{1}{4}\sum_4\sum_{j=1}^N\cos\(2\pi\[\int_0^{\abs{y}}\frac{\d\gamma}{\DD_{mn}(\gamma)} \pm \frac{\DD_{mn}^{\mathrm{rcv}}}{\DD_{mn}} \pm \frac{\DD_{mn0}^{\mathrm{src}}}{\DD_{mn0}}\]j\) \\
      \abs{B_m}^2_C &= \frac{2}{r_0r}\sum_{n=2}^\inf\frac{k_{y|mn}}{\DD_{mn}k_{r|mn}}\frac{k_{y_0|mn}}{\DD_{mn0}k_{r_0|mn}}\frac{1}{k_{y|mn}^2} \times \{\mathcal{CF}_0\} \Delta n \\
      \abs{B_m}^2 &= \frac{1}{r_0r}\sum_{n=1}^\inf\frac{k_{y|mn}}{\DD_{mn}k_{r|mn}}\frac{k_{y0|mn}}{\DD_{mn0}k_{r0|mn}}\frac{1}{k_{y|mn}^2}\Delta n \\
      &\qquad + \frac{2}{r_0r}\sum_{n=2}^\inf\frac{k_{y|mn}}{\DD_{mn}k_{r|mn}}\frac{k_{y_0|mn}}{\DD_{mn0}k_{r_0|mn}}\frac{1}{k_{y|mn}^2} \times \{\mathcal{CF}\} \Delta n \\
  \end{split}\end{equation}

  \begin{equation}\begin{split}
      &= \frac{1}{r_0r}\sum_n\frac{k_{y|mn}}{\DD_{mn}k_{r|mn}}\frac{k_{y0|mn}}{\DD_{mn0}k_{r0|mn}}\frac{1}{k_{y|mn}^2}\(1+2\{\mathcal{CF}_0\}\)\Delta n \\
      &= \frac{1}{r_0r}\int_0^{\alpha_{\mathrm{max}}}\frac{k_{y|m}}{\DD_{m}k_{r|m}}\frac{k_{y0|m}}{\DD_{m0}k_{r0|m}}\frac{1}{k_{y|m}^2}\(1+2\{\mathcal{CF}_0\}\)\dd{n}{k_{y|m}}\dd{k_{y|m}}{\alpha}\d\alpha \\
  \end{split}\end{equation}

  \begin{equation}\begin{split}
      &= \frac{1}{r_0r}\int_0^{\alpha_{\mathrm{max}}}\frac{k_{y|m}}{\DD_{m}k_{r|m}}\frac{k_{y0|m}}{\DD_{m0}k_{r0|m}}\frac{1}{k_{y|m}^2}\(1+2\{\mathcal{CF}_0\}\)\frac{\DD_m}{2\pi}k_{ry|m}\sin\alpha\d\alpha \\
      &= \frac{1}{2\pi r_0r}\int_0^{\alpha_{\mathrm{max}}}\frac{\cot(\alpha_0)}{k_{y|m}\DD_{m0}}\(1+2\{\mathcal{CF}_0\}\)\d\alpha \\
  \end{split}\end{equation}

  Now we can see clearly which terms must change in the case of range dependence in the $y$-direction.

  \begin{equation}\begin{split}
      \Theta_m(\theta) &= \sqrt{\frac{2}{\theta_B}}\sin(\vkap_m\theta) \twhere \vkap_m = (n-\half)\pi/\theta_B \\
      \left\{\mathcal{RL}\right\} &= \prod_0^{\abs{y}}\left\{\abs{\frac{k_r^<(\rint|\gamma)-k_r^>(\rint|\gamma)}{k_r^<(\rint|\gamma)+k_r^>(\rint|\gamma)}}^{2/\DD(\gamma)}\right\}^{\d\gamma} \\
      \left\{\mathcal{CF}\right\} &= \frac{1}{4}\sum_4\left\{(1+2N)\exp\[\frac{-(1+2N)^2}{\pi}\sin^2\(\pi\[\int_0^{\abs{y}}\frac{\d\gamma}{\DD_m(\gamma)}\pm\frac{\DD_m^{\mathrm{rcv}}}{\DD_m}\pm\frac{\DD_m^{\mathrm{src}}}{\DD_{m0}}\]\)\]\right\} \\
      \abs{B_m}^2 &= \frac{1}{2\pi r_0r}\int_0^{\alpha_{\mathrm{max}}}\left\{\mathcal{RL}\right\}\times\frac{\cot(\alpha_0)}{k_{y|m}\DD_{m0}}\times\left\{\mathcal{CF}\right\}\d\alpha \\
      \mathscr{P}_{\mathrm{TL}} &= S^2\sum_m\abs{\Theta_m(\theta)}^2\abs{\Theta_m(\theta_0)}^2\abs{B_m}^2 \\
      &= \sum_m\abs{\Theta_m(\theta)}^2\abs{\Theta_m(\theta_0)}^2 \left\{\frac{8\pi}{r_0r}\int_0^{\alpha_{\mathrm{max}}}\left\{\mathcal{RL}\right\}\times\frac{\cot(\alpha_0)}{k_{y|m}\DD_{m0}}\times\left\{\mathcal{CF}\right\}\d\alpha\right\} \\
      \mathrm{TL} &= -10\log_{10}\[\frac{\rho_0c_0}{\rho c}\mathscr{P}_{\mathrm{TL}}\] \\
  \end{split}\end{equation}

  \section{Implementation}

  The range-independent hybrid normal mode and energy flux model for the wedge environment with a front was implemented in Matlab.

  \begin{equation}\begin{split}
      \Theta_m(\theta) &= \sqrt{\frac{2}{\theta_B}}\sin(\vkap_m\theta) \twhere \vkap_m = (n-\half)\pi/\theta_B \\
      \left\{\mathcal{RL}\right\} &= \abs{\frac{k_r^<(\rint)-k_r^>(\rint)}{k_r^<(\rint)+k_r^>(\rint)}}^{2\abs{y}/\DD_m} \\
      \left\{\mathcal{CF}\right\} &= \frac{1}{4}\sum_4\left\{(1+2N)\exp\[\frac{-(1+2N)^2}{\pi}\sin^2\(\pi\[\pmDD\]\)\]\right\} \\
      \abs{B_m}^2 &= \frac{1}{2\pi r_0r}\int_0^{\alpha_{\mathrm{max}}}\left\{\mathcal{RL}\right\}\times\frac{\cot(\alpha_0)}{k_{y|m}\DD_m}\times\left\{\mathcal{CF}\right\}\d\alpha \\
      \mathscr{P}_{\mathrm{TL}} &= S_{\mathrm{TL}}^2\sum_m\abs{\Theta_m(\theta)}^2\abs{\Theta_m(\theta_0)}^2\abs{B_m}^2 \\
      &= \sum_m\abs{\Theta_m(\theta)}^2\abs{\Theta_m(\theta_0)}^2 \left\{\frac{8\pi}{r_0r}\int_0^{\alpha_{\mathrm{max}}}\left\{\mathcal{RL}\right\}\times\frac{\cot(\alpha_0)}{k_{y|m}\DD_m}\times\left\{\mathcal{CF}\right\}\d\alpha\right\} \\
      \mathrm{TL} &= -10\log_{10}\[\frac{\rho_0c_0}{\rho c}\mathscr{P}_{\mathrm{TL}}\] \\
  \end{split}\end{equation}

  Each angular/vertical modefunction is trivial to solve for and is orthogonal to the energy flux solution. However, the energy flux solution that represents the horizontally dependent modal coefficient must be calculated for each modenumber $m$, since the horizontal problem is dependent on $\vkap_m$ in the reduced wavenumber profile.  We also see that the WKB solutions are useful since they can deal with a radially varying wavenumber profile.

  \begin{equation}k_{ry|m}^2=k^2(r)-\frac{\vkap_m^2}{r^2}\end{equation}

  The propagation angles are defined at each receiver position and then mapped to the channel axis of maximum wavenumber, $k_{\mathrm{ax}|m}=\max(k_{ry|m})$, and to the front range as well for calculation of the reflection coefficient. Since this environment is currently range-independent in the $y$-direction, the Snell's law analog in the horizontal problem is valid between source and receiver axial range in the $y$-direction.  In an axially range dependent environment, propagation angles would be mapped from source to receiver position via the ray invariant.

  \begin{equation}\begin{split}
      k_{ry}\cos(\alpha) &= k_{ry0}\cos(\alpha_0) \\
      &= k_{ry}^{\mathrm{ax}}\cos(\alpha^{\mathrm{ax}}) \\
      &= k_{ry}^{\mathrm{frt}}\cos(\alpha^{\mathrm{frt}}) \\
  \end{split}\end{equation}

  The collection of eigenvalues, $\{\vkap_m\}$, define a set of horizontal problems to solve for the modal coefficients. The energy flux model can theoretically handle arbitrarily shaped wavenumber profiles as long as the validity conditions of the WKB approximation are not violated. Since we are considering an isospeed water column, the reduced wavenumber profile depends simply on $r^{-2}$. In this environment, every vertical mode has a shoreward turning point apex where $\vkap^2/r^2=k^2$ and the wave propagating perpendicular towards the shore reaches furthest toward the apex of the wedge. This means that every reduced wavenumber profile $k_{ry}$ will vanish at some point in radial range, every wave will refract away from the shoreline and interact with the front, and every reduced wavenumber profile can be interpolated as piecewise-linear in $r$.

  \begin{equation}\begin{split}
      k_{ry}^2 &= k^2 - \frac{\vkap^2}{r^2} \\
      \{r^{(i)},k_{ry}^{(i)}\} \tfor i &= 1,2,3,...,I  \\
      \Delta_rk_{ry}^{(i)} &= \frac{k_{ry}^{(i+1)}-k_{ry}^{(i)}}{r^{(i+1)}-r^{(i)}} \\
      k_{ry}^{(i)}(r) &= \Delta_rk_{ry}^{(i)}(r-r^{(i)}) \\
  \end{split}\end{equation}

  The cycle calculations are performed by analytically integrating each piecewise linear region of profile separately and summing them together.  The profile could be described as piecewise linear in $c$ or as piecewise linear in $k$.  Since the wavenumber profile vanishes at some point $r_m^{\mathrm{lo}}=\vkap^2/k^2$, the phase speed $c_{ry}$ necessarily diverges to infinity at this point.  Thus for numerical consistency it is preferred to perform the cycle calculations analytically with respect to the wavenumber $k_{ry}$.  Additionally, we require that the profile be singly-ducted for ease of model implementation.  It should theoretically be possible to handle multiply ducted profiles but this has implementation challenges that have been reserved as a future enhancement.

  The bulk wavenumber profile $k(r)$ as a model parameter is in this case constant and likely defined by 2 points, so we upsample the bulk wavenumber profile, calculate modal eigenvalues, calculate reduced wavenumbers using the upsampled bulk wavenumbers, and then interpolate the reduced wavenumber profile back to its vanishing point for each mode. Then for each mode and receiver position, we compute the cycle calculations exactly for each propagation angle defined at the receivers.

  \begin{equation}\begin{split}
      k_y &= k_{ry}\cos(\alpha) \\
      k_{ry}^{(i)}(r) &= \Delta_rk_{ry}^{(i)}\cdot r - \Delta_rk_{ry}^{(i)}\cdot r^{(i)} \\
      \DD &= 2\int_{r'}^{r''}\cot(\alpha)\dr \\
      &= 2\int_{r'}^{r''}\frac{k_y}{k_r}\dr \\
      &= 2\int_{r'}^{r''}\frac{1}{\sqrt{\(\frac{k_{ry}}{k_y}\)^2-1}}\dr \\
  \end{split}\end{equation}

  The Snell's law analog for the horizontal problem is used to map the propagation angles with respect to the radial range. The sound speed gradient is constant within each layer, and the cycle distance can be calculated analytically. Naturally, the upper and lower integration limits are determined by the particular trajectory of the representative ray.  A wave can either refract back when it encounters the vertexing wavenumber (a wavenumber minimum) or it does not encounter the vertexing wavenumber but instead reflects back from some boundary (our sound speed front). Regardless of the case, the integration of the cycle distance is performed exactly over either the entire layer of the sound speed profile or a partial layer due to refraction, and then the contribution from each layer is summed together.

  \begin{equation}\begin{split}
      \text{Vertexing Wavenumber:}\qquad k_y &= k_{ry0}\cos\alpha_0 = k_{ry}\cos\alpha \\
      \DD(\alpha,r) &= 2\int_{r'}^{r''}\[\(\frac{k_{ry}}{k_y}\)^2-1\]^{-\half}\dr \\
      &= 2\int_{r'}^{r''}\[\(\frac{\Delta_rk_{ry}^{(i)}(r-r^{(i)})}{k_y}\)^2-1\]^{-\half}\dr \\
      \DD^{(i)} &= 2\[\frac{k_y}{\Delta_rk_{ry}^{(i)}}\coth^{-1}\(\sqrt{1-\(\frac{k_y}{\Delta_rk_{ry}^{(i)}(r-r^{(i)})}\)^2}\)\]_{r'}^{r''} + \CC \\
      &= 2\[\frac{k_y}{\Delta_rk_{ry}^{(i)}}\coth^{-1}\(\sqrt{1-\(\frac{k_y}{k_{ry}^{(i)}(r)}\)^2}\)\]_{r'}^{r''} + \CC \\
  \end{split}\end{equation}

  The partial cycle distances are also calculated for determination of the convergence factor. These are evaluated from an arbitrary but consistent reference radial range $r_{\mathrm{ref}}$ to the queried radial range, either $r$ at the receiver position or $r_0$ at the source position \cite{harrison2013ray}.

  \begin{equation}\begin{split}
      \DD_m^{\mathrm{rcv}} &= \int_{r_{\mathrm{ref}}}^r \cot \alpha \dr \\
      \DD_m^{\mathrm{src}} &= \int_{r_{\mathrm{ref}}}^{r_0} \cot\alpha\dr \\
  \end{split}\end{equation}

  The value of the radial wavenumber at the front is used to calculate a propagation angle dependent reflection coefficient for the frontal interface.  This reflection loss is applied gradually over a cycle distance.

  \begin{equation}\begin{split}
      \{\mathcal{RL}\} &= \abs{\frac{k_r^<(\rint)-k_r^>(\rint)}{k_r^<(\rint)+k_r^>(\rint)}}^{2\abs{y}/\DD}
  \end{split}\end{equation}

  The convergence factor is straightforward to compute once we have the complete and partial cycle distance calculations. The maximum mode number difference, $N$, is a parameter that focuses the peaks of the approximate interference function.  The function inside of the fourfold summation approximates the interference between the four ray families, becoming non-zero when a ray family cycle is completed (a whole cycle distance is traversed) including the cycle phase offsets for the source and receiver positions \cite{harrison2013ray}.

  \begin{equation}\begin{split}
      \{\mathcal{CF}\} &= \frac{1}{4}\sum_4\left\{(1+2N)\exp\[\frac{-(1+2N)^2}{\pi}\sin^2\(\pi\[\frac{\abs{y}}{\DD_m}\pm\frac{\DD_m^{\mathrm{src}}}{\DD_m}\pm\frac{\DD_m^{\mathrm{rcv}}}{\DD_m}\]\)\]\right\} \\
  \end{split}\end{equation}

  With all of the integrand factors, we integrate over the propagation angle using composite Simpson's rule. The angular modefunctions are evaluated at the source and receiver positions, the semi-coherent pressure squared is summed over all modes, and this is done for each receiver position. Lastly, the transmission loss is calculated by the log of the product of the characteric acoustic impedance ratio between source and receiver positions and the semi-coherent pressure squared.

  \begin{equation}\begin{split}
      \mathscr{P}_{\mathrm{TL}} &= \sum_m\abs{\Theta_m(\theta)}^2\abs{\Theta_m(\theta_0)}^2 \left\{\frac{8\pi}{r_0r}\int_0^{\alpha_{\mathrm{max}}}\left\{\mathcal{RL}\right\}\times\frac{\cot(\alpha_0)}{k_{y|m}\DD_m}\times\left\{\mathcal{CF}\right\}\d\alpha\right\} \\
      \mathrm{TL} &= -10\log_{10}\[\frac{\rho_0c_0}{\rho c}\mathscr{P}_{\mathrm{TL}}\] \\
  \end{split}\end{equation}

  \section{Model Comparison}

  The test environment and scenario we are feeding into our model matches the one specified in ``Analytical study of the horizontal ducting of sound by an oceanic front over a slope'' \cite{lin2012analytical}.  The source frequency was at $25\unit{Hz}$ which is fairly low for the energy flux model.  The energy flux model relies on the validity of the continuum of modes assumption, which boils down to a high frequency argument comparing the size of the duct-transecting wavelength to the duct width.  This can be reinterpreted as a requirement for the number of propagating modes (low loss, trapped modes, beneath the critical angle) \cite{holland2010propagation}.  We only need to consider an ideal waveguide to see this relationship.  In an isovelocity horizontally bounded waveguide with homogeneous Dirichlet and Neumann boundary conditions, we will assume the existence of a critical angle though we typically need a fluid acoustic interface for a critical angle of reflection to exist.  We wish to ensure that the steepest propagating mode has many nodes.  We use the closed phase integral to count the number of propagating modes.

  \begin{equation}\begin{split}
      \int k_z \dz &= (n-\half)\pi \\
      \int k_{z,c} \dz &\gg 1 \\
      \int k\sin\theta_c \dz &\gg 1 \\
      kH\sin\theta_c &\gg 1 \\
  \end{split}\end{equation}

\begin{figure}[ht]
    \includegraphics[width=\columnwidth]{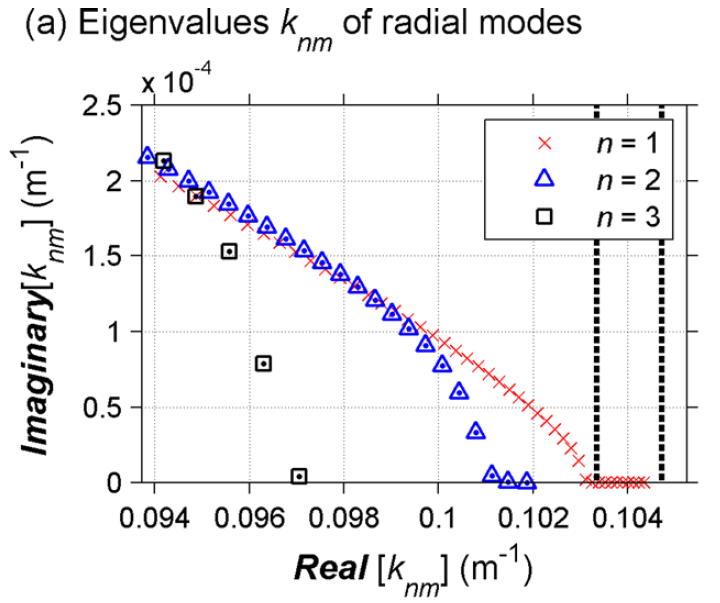}
    \caption{\label{fig:eigenvalues}{Plot of eigenvalues for angular (vertical) modes 1 2 and 3. Adapted from Lin and Lynch 2012 \cite{lin2012analytical}}}
  \end{figure}

  This relation is analogous in our model with the exception of working with different coordinate geometries. Lin and Lynch 2012 provides a figure plotting the eigenvalues at $25\unit{Hz}$ source frequency, which we have adapted here in Fig.[\ref{fig:eigenvalues}].

  From Fig.[\ref{fig:eigenvalues}], we see that only angular (vertical) mode 1 has enough propagating modes to make an adequate comparison to the energy flux model.  The propagating modes (trapped modes) are those with small imaginary wavenumber components meaning they experience very little loss as they propagate in the $y$-direction. Angular mode 2 has only two modes that should be considered trapped which means that it would make for a poor comparison to a continuum of modes model.  For this reason we will primarily compare results for mode 1.

  The section of the TL field we are considering is a constant-$\theta$ slice at the same depth as the source, i.e. $\theta_{\mathrm{rcv}}=1.5^{\circ}$.  The 2D transmission loss plot extends in both the $r$ and $y$ directions, with the bottom axis corresponding to the radial range of the sound velocity front.

  \begin{figure}[ht]
    \includegraphics[width=\columnwidth]{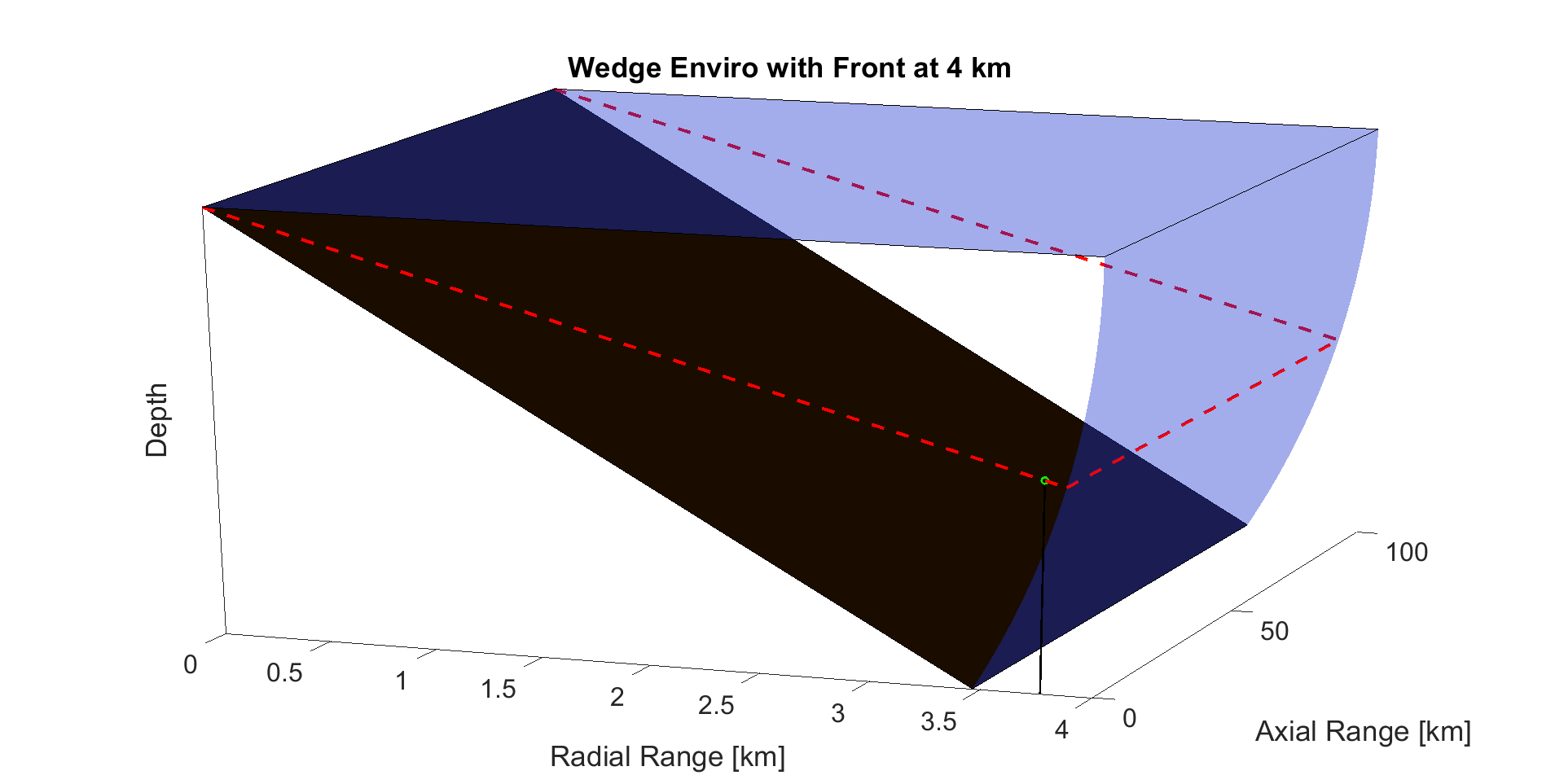}
    \caption{\label{fig:sliceOfWedge}{Wedge environment with constant $\theta$ slice outlined by the red dashed line.}}
  \end{figure}

  First we show a complete comparison of the sound field, followed by a closer comparison of the transmission losses for mode 1.  Since the normal mode model shows much more interference structure than the energy flux model, a uniform averaging window is applied to the normal mode intensity with dimensions of $\Delta y = 2\unit{km}$ and $\Delta r = 0.2\unit{km}$.

  \begin{figure}[ht]
    \includegraphics[width=\columnwidth]{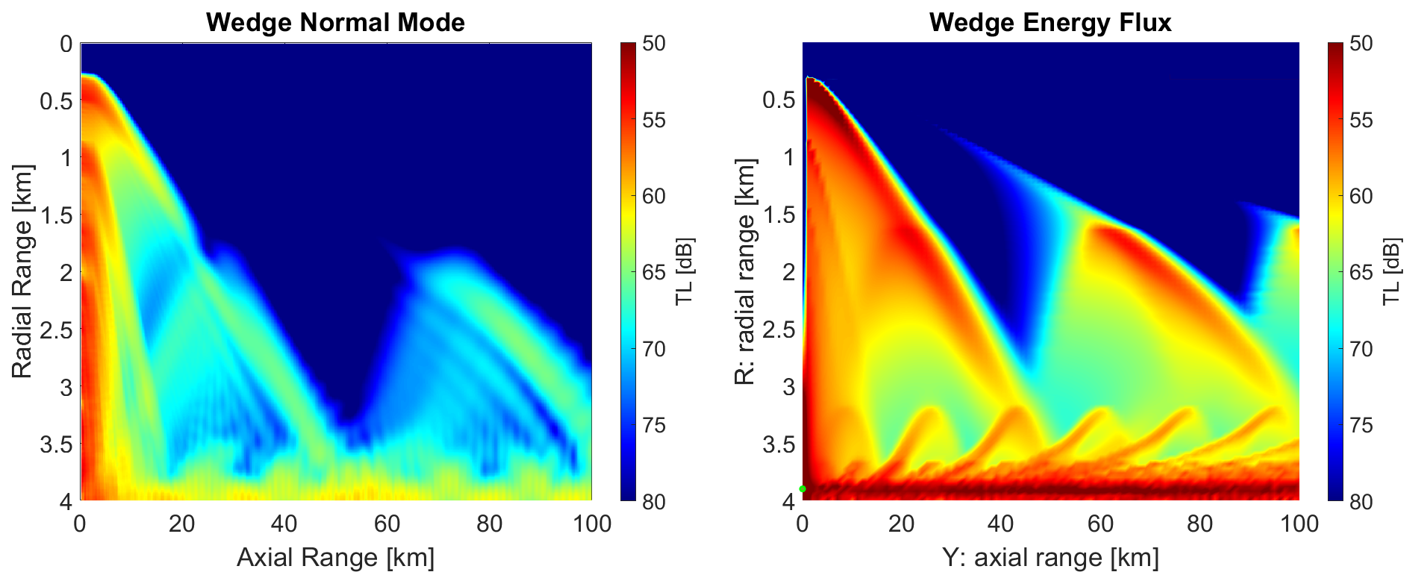}
    \caption{\label{fig:tlAllModes}{Comparison of Transmission Loss for all angular modes between the normal mode model (left) and the energy flux model (right)}}
  \end{figure}

  \begin{figure}[ht]
    \includegraphics[width=\columnwidth]{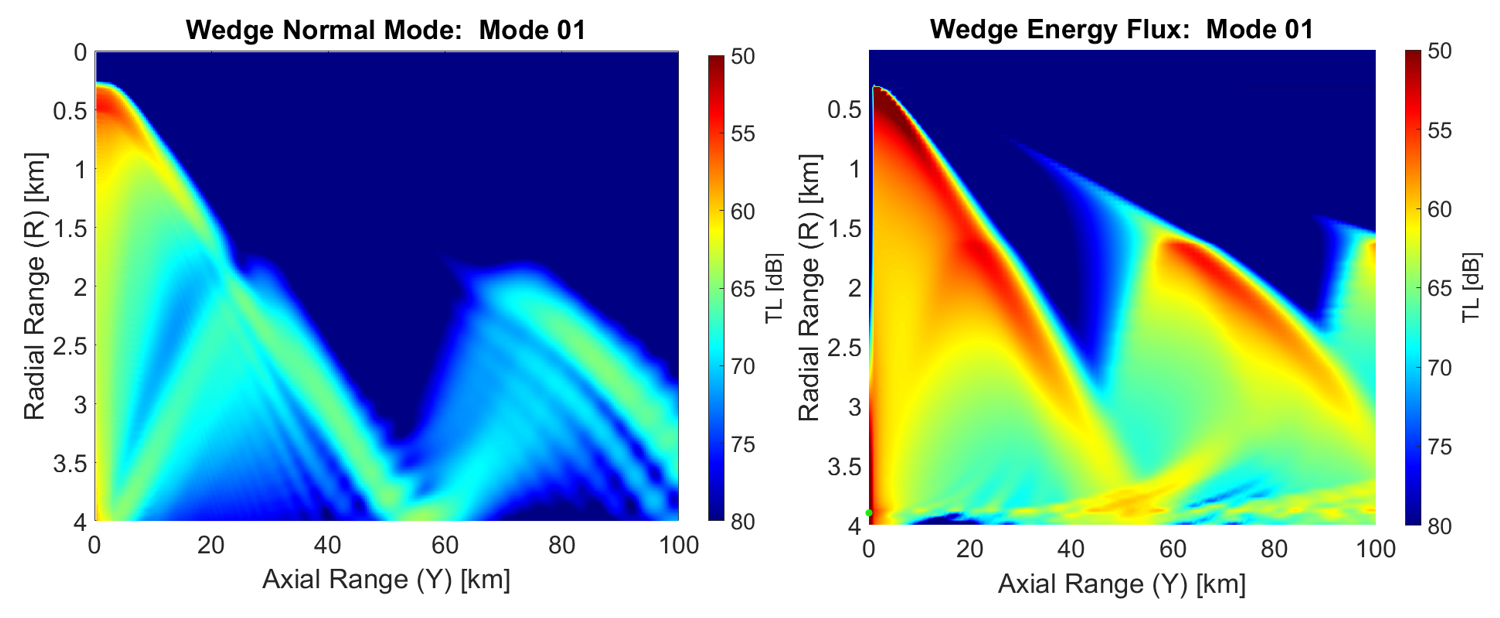}
    \caption{\label{fig:tlAllModes}{Comparison of Transmission Loss for only angular mode 01 between the normal mode model (left) and the energy flux model (right)}}
  \end{figure}

  From the transmission loss plots, we see general agreement in the shape and location of the convergent field structures.  The leaky radial modes for angular mode 1 (those with reflection loss at the front) create the bell-shaped structure near the source axial range.  At further ranges, the leaky radial modes have decayed and the propagating radial modes for angular mode 1 form an interference pattern which is captured by the energy flux solution.

  However there is currently a discrepancy between the overall transmission loss between these models.  Possible sources include analytical error in the derivation of the model, violation of an assumption or approximation, or inconsistencies with mathematical and physical conventions.

  \begin{figure}[ht]
    \includegraphics[width=\columnwidth]{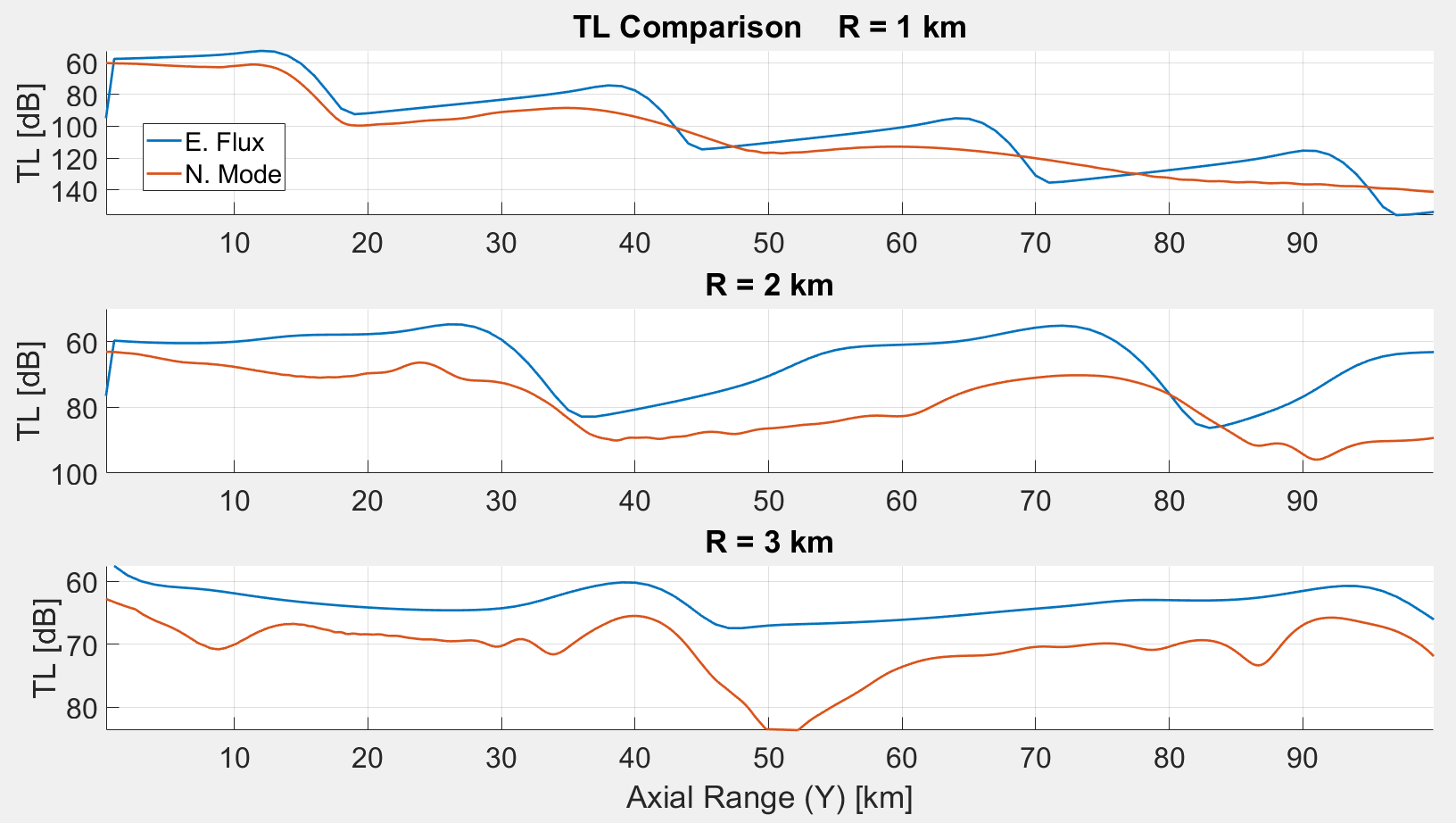}
    \caption{\label{fig:rSlices}{Comparison of Transmission Loss for only angular mode 01, constant axial range.}}
  \end{figure}

  \begin{figure}[ht]
    \includegraphics[width=\columnwidth]{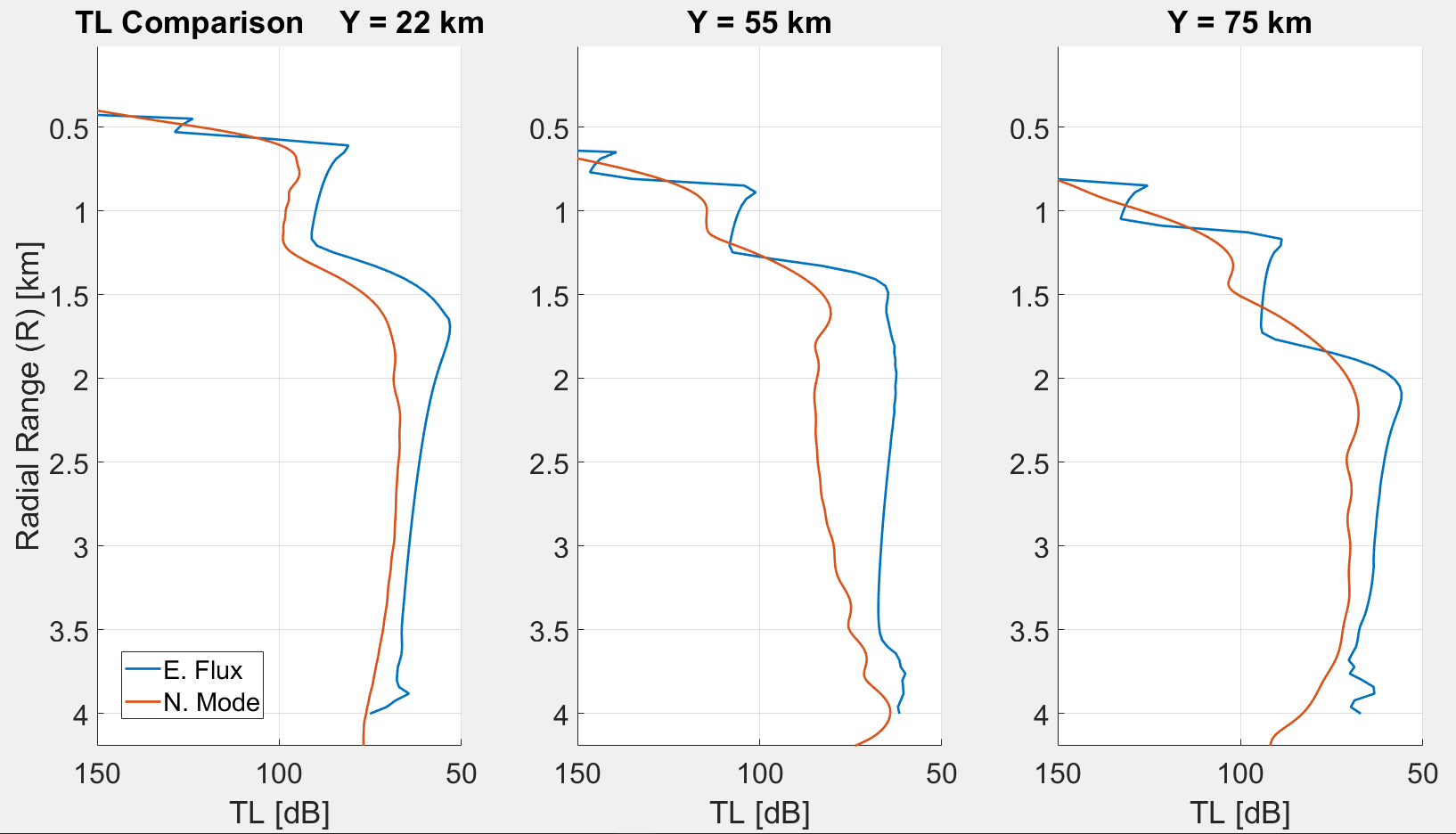}
    \caption{\label{fig:ySlices}{Comparison of Transmission Loss for only angular mode 01, constant radial range.}}
  \end{figure}

  We have looked at the reflection coefficient calculations as a potential source of error.  Lin and Lynch assumed Bessel function radial modes and solved for the reflection coefficient by applying the smoothness boundary conditions at the frontal interface \cite{lin2012analytical}.  In this paper we have assumed WKB (complex exponential) radial modes and then applied that smoothness boundary conditions at the front interface. Upon comparison of the two methods, the reflection coefficients are nearly identical for the first mode when the eigenvalues are assumed real, so it seems unlikely that this explains the TL discrepancy.

  We have also looked at the source monopole amplitude and the definition of the transmission loss. Both models should theoretically be in agreement on these points since both models seem to assume a monopole source amplitude of $4\pi$, which corresponds to a reference intensity level of $1\unit{W/m^2}$.

  \clearpage

  \section{Conclusion}
  We were able to show general agreement in the location and shape of the caustic features for angular mode 1. The radial modal density for mode 1 is sufficient for a comparison to the energy flux model, but the higher-order angular modes are too sparse in radial modes at this frequency.  Both of these models can be run at a higher frequency which may provide a better comparison, especially for the higher order modes.

  The WKB modes have amplitudes that diverge to infinity as the waves approach their turning points.  Chris Harrison has shown in the 2D model that the WKB mode amplitudes can be limited (capped off) as a sort of crude approximation to the Airy function used to stitch together the oscillatory and evanescent regions of WKB modes \cite{harrison2013ray}.  In this model, an Airy function would not be an appropriate turning point solution to perform asymptotic matching of the oscillatory and evanescent WKB mode functions.  It should be possible to derive a limiting ceiling value based on the amplitude peak of the Bessel function solutions, i.e. instead of approximating linear sound speed in the vicinity of the turning point, approximate with a constant sound speed and then the solution will be known to be a superposition of Bessel functions, the amplitude of which should have a maximum in the vicinity of the turning point.

  This model also does not consider interference between the angular modes; in fact it only considers the interference between the radial modes for the coherent intensity summation of vertical modes.  It is possible to incorporate further interference effects, however the method used in this paper follows closest to the convergence factor recently derived \cite{harrison2013ray}.

  At the end of the derivation section, we showed how to add axial range dependence for the convergence factor using the adiabatic modes approximation \cite{harrison2015efficient}.  The method is directly analogous in this model.  Range-dependence in this model was not implemented since it adds algorithmic complexity and we intend to work on a more generalized 3D energy flux model in the near future.

  This paper outlines one way of incorporating the energy flux method into a 3D ocean acoustic propagation model.  However, this model is rather limited in the environments that it can apply to.  It might be theoretically possible to generalize to a more complicated geometry in this coordinate system, but it is less than ideal.  Using similar theoretical tools, we would like to develop a generalized 3D semi-coherent energy flux model that is based on a Cartesian coordinate system and does not assume azimuthal symmetry.

  \section*{Acknowledgments}
  This research was supported by The Office of Naval Research under the NDSEG fellowship program.

  \bibliographystyle{unsrt}
  \bibliography{bibliography.bib}

\end{document}

%% file: definitions.tex
\renewcommand{\(}{\left(}
\renewcommand{\)}{\right)}
\renewcommand{\[}{\left[}
\renewcommand{\]}{\right]}

\renewcommand{\deg}{\degree}
\renewcommand{\vec}[1]{\overrightarrow{#1}}
\newcommand{\bvec}[1]{\bm{#1}}
\newcommand{\xdot}{\dot{x}}
\newcommand{\ydot}{\dot{y}}
\newcommand{\zdot}{\dot{z}}
\newcommand{\xhat}{\hat{x}}
\newcommand{\yhat}{\hat{y}}
\newcommand{\zhat}{\hat{z}}
\newcommand{\ihat}{\hat{i}}
\newcommand{\jhat}{\hat{j}}
\newcommand{\khat}{\hat{k}}
\newcommand{\ex}{\vec{e}_x}
\newcommand{\ey}{\vec{e}_y}
\newcommand{\ez}{\vec{e}_z}
\newcommand{\grad}{\vec{\nabla}}
\renewcommand{\div}{\grad\cdot}
\newcommand{\curl}{\grad\times}
\newcommand{\lap}{\nabla^2}
\newcommand{\bgrad}{\bvec{\nabla}}
\newcommand{\bdiv}{\bgrad\cdot}
\newcommand{\bcurl}{\bgrad\times}
\newcommand{\block}[2]{\thead{\text{#1}\\\text{#2}}}
\newcommand{\abs}[1]{\left|#1\right|}
\newcommand{\norm}[1]{\left\lVert#1\right\rVert}
\newcommand{\ee}[1]{\cdot10^{#1}}
\newcommand{\evalat}[2]{\left.#1\right|_{#2}}
\newcommand{\absevalat}[2]{\left|#1\right|_{#2}}
\newcommand{\evalover}[3]{\left.#1\right|_{#2}^{#3}}
\newcommand{\bevalat}[2]{\left[#1\right]_{#2}}
\newcommand{\bevalover}[3]{\left[#1\right]_{#2}^{#3}}
\renewcommand{\Re}[1]{\mathfrak{Re}\left\{#1\right\}}
\renewcommand{\Im}[1]{\mathfrak{Im}\left\{#1\right\}}
\newcommand{\eps}{\epsilon}
\newcommand{\veps}{\varepsilon}
\newcommand{\vrho}{\varrho}
\newcommand{\vkap}{\varkappa}
\newcommand{\vthe}{\vartheta}
\newcommand{\vpi}{\varpi}
\newcommand{\vsig}{\varsigma}
\newcommand{\vphi}{\varphi}
\newcommand{\laro}{\leftarrow}
\newcommand{\Laro}{\Leftarrow}
\newcommand{\raro}{\rightarrow}
\newcommand{\Raro}{\Rightarrow}
\newcommand{\supm}[1]{\underset{#1}{\mathrm{sup}}}
\newcommand{\infm}[1]{\underset{#1}{\mathrm{inf}}}
\renewcommand{\inf}{\infty}
\renewcommand{\lim}[1]{\underset{#1}{\mathrm{lim}}\;}
\newcommand{\Ai}{\mathrm{Ai}}
\newcommand{\Bi}{\mathrm{Bi}}
\newcommand{\floor}[1]{\left\lfloor #1 \right\rfloor }
\newcommand{\ceil}[1]{\left\lceil #1 \right\rceil }
\newcommand{\wtil}[1]{\widetilde{#1}}
\newcommand{\bra}[1]{\left<#1\right|}
\newcommand{\ket}[1]{\left|#1\right>}
\newcommand{\braket}[2]{\left<#1\middle|#2\right>}
\newcommand{\ngtv}{\scalebox{0.75}[1.0]{$^-$}}
\newcommand{\sgn}{\mathrm{sgn}}

\newcommand{\D}{\mathrm{D}}
\newcommand{\p}{\hspace{.2777em}\partial}
\newcommand{\pp}[2]{\frac{\p #1}{\p #2}}
\newcommand{\pnp}[3]{\frac{\p^{#1} #2}{\p {#3}^{#1}}}
\renewcommand{\d}{\hspace{.2777em}\mathrm{d}}
\newcommand{\dd}[2]{\frac{\d #1}{\d #2}}
\newcommand{\dnd}[3]{\frac{\d^{#1} #2}{\d {#3}^{#1}}}
\newcommand{\da}{\d a}
\newcommand{\db}{\d b}
\newcommand{\dc}{\d c}
\newcommand{\de}{\d e}
\newcommand{\df}{\d f}
\newcommand{\dg}{\d g}
\newcommand{\di}{\d i}
\newcommand{\dk}{\d k}
\newcommand{\dl}{\d l}
\newcommand{\dm}{\d m}
\newcommand{\dn}{\d n}
\newcommand{\dq}{\d q}
\newcommand{\dr}{\d r}
\newcommand{\ds}{\d s}
\newcommand{\dt}{\d t}
\newcommand{\du}{\d u}
\newcommand{\dv}{\d v}
\newcommand{\dw}{\d w}
\newcommand{\dx}{\d x}
\newcommand{\dy}{\d y}
\newcommand{\dz}{\d z}
\newcommand{\dA}{\d A}
\newcommand{\dB}{\d B}
\newcommand{\dC}{\d C}
\newcommand{\dD}{\d D}
\newcommand{\dE}{\d E}
\newcommand{\dF}{\d F}
\newcommand{\dG}{\d G}
\newcommand{\dH}{\d H}
\newcommand{\dI}{\d I}
\newcommand{\dJ}{\d J}
\newcommand{\dK}{\d K}
\newcommand{\dL}{\d L}
\newcommand{\dM}{\d M}
\newcommand{\dN}{\d N}
\newcommand{\dO}{\d O}
\newcommand{\dP}{\d P}
\newcommand{\dQ}{\d Q}
\newcommand{\dR}{\d R}
\newcommand{\dS}{\d S}
\newcommand{\dT}{\d T}
\newcommand{\dU}{\d U}
\newcommand{\dV}{\d V}
\newcommand{\dW}{\d W}
\newcommand{\dX}{\d X}
\newcommand{\dY}{\d Y}
\newcommand{\dZ}{\d Z}
\newcommand{\dalpha}{\d\alpha}
\newcommand{\dbeta}{\d\beta}
\newcommand{\dgamma}{\d\gamma}
\newcommand{\ddelta}{\d\delta}
\newcommand{\depsilon}{\d\epsilon}
\newcommand{\dzeta}{\d\zeta}
\newcommand{\deta}{\d\eta}
\newcommand{\dtheta}{\d\theta}
\newcommand{\diota}{\d\iota}
\newcommand{\dkappa}{\d\kappa}
\newcommand{\dlambda}{\d\lambda}
\newcommand{\dmu}{\d\mu}
\newcommand{\dnu}{\d\nu}
\newcommand{\dxi}{\d\xi}
\newcommand{\domicron}{\d\omicron}
\newcommand{\dpi}{\d\pi}
\newcommand{\drho}{\d\rho}
\newcommand{\dsigma}{\d\sigma}
\newcommand{\dtau}{\d\tau}
\newcommand{\dupsilon}{\d\upsilon}
\newcommand{\dphi}{\d\phi}
\newcommand{\dchi}{\d\chi}
\newcommand{\dpsi}{\d\psi}
\newcommand{\domega}{\d\omega}

\newcommand{\Let}{\text{Let:}\qquad}
\newcommand{\Def}{\text{Def:}\qquad}
\newcommand{\BCs}{\text{BCs:}\qquad}
\newcommand{\NB}{\text{NB:}\qquad}
\newcommand{\tif}{\qquad\text{if}\qquad}
\newcommand{\tiff}{\qquad\text{iff}\qquad}
\newcommand{\tthen}{\qquad\text{then}\qquad}
\newcommand{\Then}{\text{Then}\qquad}
\newcommand{\tis}{\qquad\text{is}\qquad}
\newcommand{\tand}{\qquad\text{and}\qquad}
\newcommand{\tor}{\qquad\text{or}\qquad}
\newcommand{\tfor}{\qquad\text{for}\qquad}
\newcommand{\twhere}{\qquad\text{where}\qquad}
\newcommand{\Where}{\text{Where:}\qquad}
\newcommand{\Ansatz}{\text{Ansatz:}\qquad}
\newcommand{\Assume}{\text{Assume:}\qquad}
\newcommand{\Recall}{\text{Recall:}\qquad}
\newcommand{\Thus}{\text{Thus:}\qquad}
\newcommand{\twhile}{\qquad\text{while}\qquad}

\newcommand{\II}{\mathbb{I}}
\newcommand{\DD}{\mathbb{D}}
\newcommand{\TT}{\mathbb{T}}
\newcommand{\CC}{\mathbb{C}}
\newcommand{\RR}{\mathbb{R}}

\newcommand{\half}{\nicefrac{1}{2}}
\newcommand{\threehalf}{\nicefrac{3}{2}}
\newcommand{\fivehalf}{\nicefrac{5}{2}}
\newcommand{\third}{\nicefrac{1}{3}}
\newcommand{\twothird}{\nicefrac{2}{3}}
\newcommand{\quarter}{\nicefrac{1}{4}}
\newcommand{\threequarter}{\nicefrac{3}{4}}
\newcommand{\fivequarter}{\nicefrac{5}{4}}
\newcommand{\eighth}{\nicefrac{1}{8}}
\newcommand{\threeeighth}{\nicefrac{3}{8}}
\newcommand{\fiveeighth}{\nicefrac{5}{8}}
\newcommand{\sixteenth}{\nicefrac{1}{16}}
\newcommand{\threesixteenth}{\nicefrac{3}{16}}
\newcommand{\fivesixteenth}{\nicefrac{5}{16}}